\documentstyle[epsfig,aps,pre,multicol]{revtex}

\newcommand{\kb}{{k_{\rm B}}}
\newcommand{\ljfrq}{{\,(\epsilon/m\sigma^2)^{1/2}}}
\newcommand{\ljtime}{{\,(m\sigma^2/\epsilon)^{1/2}}}
\newcommand{\mfrq}{{\,(\epsilon/mr_{\rm e}^2)^{1/2}}}
\newcommand{\mtime}{{\,(mr_{\rm e}^2/\epsilon)^{1/2}}}

\begin{document}

\draft

\title{Structural relaxation in atomic clusters: Master equation dynamics}
\author{Mark A.~Miller, Jonathan P.~K.~Doye and David J.~Wales\thanks{Corresponding author}}
\address{University Chemical Laboratories, Lensfield Road, Cambridge CB2 1EW, UK}
\maketitle

\begin{abstract}
The role of the potential energy landscape in determining the relaxation dynamics
of model clusters is studied using a master equation. Two types of energy landscape
are examined: a single funnel, as exemplified by 13-atom Morse clusters, and
the double funnel landscape of the 38-atom Lennard-Jones cluster. Interwell rate
constants are calculated using Rice--Ramsperger--Kassel--Marcus theory within the
harmonic approximation, but anharmonic model partition functions are also considered.
Decreasing the range of the potential in the Morse clusters is shown to hinder relaxation
towards the global minimum, and this effect is related to the concomitant changes in
the energy landscape. The relaxation modes that emerge from the master equation are
interpreted and analysed to extract interfunnel rate constants for the Lennard-Jones
cluster. Since this system is too large for a complete characterization of the
energy landscape, the conditions under which the master equation can be applied to
a limited database are explored. Connections are made to relaxation processes in proteins
and structural glasses.
\end{abstract}


\begin{multicols}{2}

\section{Introduction}

Some of the most interesting processes in chemical physics involve relaxation from a
non-equilibrium state. Examples include the folding of a protein from a denatured conformation
and the formation of a crystal or glass upon cooling a liquid. Given a method for calculating
the rate constants for processes between mutually accessible states, the evolution of a
non-equilibrium probability distribution can be described by a master equation\cite{Kampen81a}.
\par
A natural way to define a state in the master equation is provided by an ``inherent
structure'' analysis of the potential energy surface (PES)\cite{Stillinger82a}.
Except at high temperatures, the configuration of an interacting system oscillates in
the basin of attraction surrounding a local minimum on the PES, and sporadically undergoes
transitions into neighbouring basins of attraction. A local minimum can therefore be
regarded as a single state in the master equation, and transition states on the PES provide
the means for dynamics to occur between the minima. We have recently obtained databases
of minima and transition states for a variety of systems\cite{Wales98c,Miller99a,LJpaper,p46},
providing the necessary ingredients for a master equation study.
There are at least two advantages to modelling relaxation in this coarse-grained
state-to-state way. Firstly, the master equation can usually be solved for much longer
time scales than are accessible by direct simulations in which the equations of motion
are integrated. Secondly, the master equation describes the relaxation of an ensemble
without the need for explicit averaging over separate trajectories. In fact, the master
equation can be used as a guide for simulations, for example to devise optimal annealing
schedules\cite{Kunz98a}.
\par
In this contribution, the master equation is applied to structural databases that we have
previously derived for some atomic clusters. Section \ref{sect:methods} summarizes the master
equation technique and methods for obtaining state-to-state rate constants in the
microcanonical and canonical ensembles. Section \ref{sect:morse} presents results for
13-atoms Morse clusters, M$_{13}$. The energy landscapes of these clusters each resemble
a funnel, in which the minima are organized into pathways of decreasing energy that lead
to the global minimum on the PES. The characteristics of the funnel change with the
range of the potential, and have been studied in detail in previous work\cite{Miller99a}.
Section \ref{sect:morse} also includes a discussion of harmonic and anharmonic partition
function models for describing equilibrium properties of the clusters. In
Sec.\ \ref{sect:thirtyeight}, dynamics on the paradigmatic double-funnel energy landscape
of the 38-atom Lennard-Jones cluster, LJ$_{38}$, are studied. We have previously made a number
of predictions concerning the dynamics of M$_{13}$\cite{Miller99a} and LJ$_{38}$\cite{LJpaper},
which can now be examined. We will also present some new ways to interpret solutions of the master
equation and extract information from them. The conditions under which the master equation treatment
of interwell dynamics is valid will also be addressed, especially in the case of LJ$_{38}$, where
the knowledge of the energy landscape is incomplete. Finally, Sec.\ \ref{sect:summary}
summarizes the main conclusions from this work.

\section{Methods}
\label{sect:methods}

\subsection{The Master Equation}
\label{sect:master}

Let ${\bf P}(t)$ be a vector whose components, $P_i(t)$ ($1\le i\le n_{\rm min}$),
are the probabilities of the cluster residing in a potential well of the geometrical
isomer $i$ at time $t$, the total number of such isomers being $n_{\rm min}$.
The time evolution of these probabilities is governed by
\begin{equation}
\frac{dP_i(t)}{dt}=\sum_{j\ne i}^{n_{\rm min}}[k_{ij}P_j(t)-k_{ji}P_i(t)],
\label{eq:dpdt}
\end{equation}
where $k_{ij}$ is the first order rate constant for transitions from well
$j$ to well $i$. 
We can set up a transition matrix matrix, $\bf W$, with components
\begin{equation}
W_{ij}=k_{ij}-\delta_{ij}\sum_{m=1}^{n_{\rm min}} k_{mi},
\end{equation}
so that the diagonal components $W_{ii}$ contain minus the total rate constant
out of minimum $i$. This definition allows us to write the set of coupled linear
differential equations (\ref{eq:dpdt})---the ``master equation''---in matrix form:
\begin{equation}
\frac{d{\bf P}(t)}{dt} = {\bf WP}(t).
\label{eq:master}
\end{equation}
If $\bf W$ cannot be decomposed into block form then the system has a uniquely
defined equilibrium state, ${\bf P}^{\rm eq}$, for which
$(d{\bf P}/dt)|_{{\bf P}={\bf P}^{\rm eq}}=0$,
i.e., ${\bf W}$ has a single zero eigenvalue whose eigenvector is the equilibrium
probability distribution. $\bf W$ is asymmetric, but can be symmetrized using
the condition of detailed balance: at equilibrium,
\begin{equation}
W_{ij}P^{\rm eq}_j = W_{ji}P^{\rm eq}_i,
\label{eq:detbal}
\end{equation}
so that $\tilde W_{ij}=(P^{\rm eq}_j/P^{\rm eq}_i)^{1/2} W_{ij}$ is symmetric.
$\bf W$ and $\tilde{\bf W}$ have the same eigenvalues, $\lambda_i$, and their
respective normalized eigenvectors ${\bf u}^{(i)}$ and $\tilde{\bf u}^{(i)}$ are related
by ${\bf u}^{(i)}={\bf S}\tilde{\bf u}^{(i)}$, where $\bf S$ is the diagonal matrix
$S_{ii}=\sqrt{P^{\rm eq}_i}$. Hence, individual components of the
eigenvectors are related by $u^{(i)}_j=\tilde u^{(i)}_j\sqrt{P^{\rm eq}_j}$.
The solution of Eq.\ (\ref{eq:master}) is then\cite{Kampen81a,Czerminski90a}
\begin{equation}
P_i(t)=\sqrt{P^{\rm eq}_i}\sum_{j=1}^{n_{\rm min}}{\tilde u}_i^{(j)} e^{\lambda_j t}
\left[\sum_{m=1}^{n_{\rm min}}{\tilde u}_m^{(j)}\frac{P_m(0)}{\sqrt{P^{\rm eq}_m}}\right],
\label{eq:msolution}
\end{equation}
where ${\tilde u}_m^{(j)}$ is component $m$ of $\tilde{\bf u}^{(j)}$.
\par
Apart from the zero eigenvalue, all the $\lambda_j$ are negative\cite{Kampen81a}.
We label eigenvalues and eigenvectors in order of decreasing
algebraic value of the eigenvalue, so that $\lambda_1=0$, and $\lambda_j<0$ for
$2\le j\le n_{\rm min}$. As $t\to\infty$, only the
$j=1$ term in Eq.\ (\ref{eq:msolution}) survives, and ${\bf P}(t)\to{\bf P}^{\rm eq}$.
This limit defines the baseline to which the remaining modes decay exponentially.
The size of the contribution of mode $j$ to the evolution of the probability of
minimum $i$ depends on component $i$ of eigenvector $j$, and on a weighted
overlap between the initial probability vector and eigenvector $j$, i.e., the
term in square brackets in Eq.\ (\ref{eq:msolution}). The sign of the product of these
two quantities determines whether the mode makes an increasing or decreasing
contribution with time.
Combinations of modes with different signs give rise to the possibility
of the accumulation and subsequent decay of transient populations as probability
flows from the initial state to equilibrium via intermediates.
\par
Eq.\ (\ref{eq:msolution}) requires the diagonalisation of the matrix $\tilde{\bf W}$,
whose dimension is the number of minima in the database, $n_{\rm min}$. The time
required to compute the eigenvectors scales as the cube of the dimension,
and the storage requirements scale as its square. (Although $\tilde{\bf W}$
may be sparse, its eigenvectors are not.) However, once diagonalisation has
been achieved, $P_i(t)$ can be calculated independently for any minimum $i$
at any instant $t$. The only restriction on $t$ comes from the accuracy
to which the eigenvalues, $\lambda_i$, can be obtained; if the error is
of the order $\delta\lambda$, Eq.\ (\ref{eq:msolution}) may diverge as $t$ approaches
$1/\delta\lambda$.
\par
An alternative way of solving the master equation is to integrate Eq.\ (\ref{eq:master})
numerically. This approach has the advantage of not requiring diagonalisation of
$\tilde{\bf W}$, and is therefore the only way to proceed for large databases.
However, it has a number of disadvantages. Firstly,
knowledge of the eigenvalues and eigenvectors of $\tilde{\bf W}$
is useful in interpreting the time evolution of ${\bf P}(t)$.
Secondly, accurate integration over long periods can be very
slow, since the accumulation of numerical error can cause the sum of the
probabilities to diverge rapidly. Thirdly, the full probability
vector ${\bf P}(t)$ (rather than selected components) must be propagated, and
the integration must start from the time at which the initial probabilities
are specified.
\par
If the initial probability vector, ${\bf P}(0)$, is strongly
non-equilibrium, many components of ${\bf P}(t)$ change rapidly as soon as
the integration starts, and then relax more slowly towards equilibrium. Therefore,
when numerically integrating the master equation, the step size required for
a given accuracy is usually smaller when $t$ is closer to zero, and
can be enlarged as $t$ grows. To take
advantage of this, the numerical integration in the present work was performed
using a Bulirsch--Stoer algorithm with an adaptive step size\cite{Press92a}.
Results from this method coincided precisely with those of the analytic solution,
where the latter could be determined.
\par
The linearity of the master equation rests on the assumption that the underlying
dynamics are Markovian. The probability of the transition $i\to j$ must not depend
on the history of reaching minimum $i$,
so that the elements of the transition matrix are indeed constants for a given
temperature or total energy. For this restriction to apply, states within a potential
well must equilibrate on a time scale faster than transitions to different
minima, so that the transitions are truly stochastic. Previous results for other
clusters\cite{Vekhter97a,Miller97a} suggest that intrawell equilibration is quite rapid. 
The Markovian requirement will impose an upper limit to the temperatures at which the master
equation can be applied to transitions between minima, since at high temperatures the
phase point does not reside in any one well long enough to establish equilibrium
within it. Division of configuration space into the basins of attraction surrounding
the minima is less useful in this dynamical regime.

\subsection{Rate Constants and Equilibrium Properties}
\label{sect:rateseqm}

To model the probability flow within a database of minima using the master equation
requires knowledge of the rate constants, $k_{ij}$, for their interconversion.
For the path from minimum $j$ through a particular transition state, denoted $\dag$,
a general form for the rate constant is provided by RRKM theory\cite{Gilbert90a}
\begin{equation}
k^\dag_j(E)=\frac{W^\dag(E)}{h\Omega_j(E)}.
\label{eq:RRKM}
\end{equation}
Here, $\Omega_j(E)$ is the density of states associated with minimum $j$, and
\begin{equation}
W^\dag(E)=\int_{V^\dag}^E \Omega^\dag(E')dE'
\label{eq:wdag}
\end{equation}
is the sum of states at the transition state with the reactive mode removed;
$\Omega^\dag (E)$ is the density of states at the transition state excluding
this mode, and $V^\dag$ is the potential energy of the transition state.
The total rate constant, $k_{ij}$, for the process $j\to i$, is then
obtained by summing Eq.\ (\ref{eq:RRKM}) over all transition states linking $j$ and $i$.
\par
Rates in the canonical ensemble, i.e., as a function of temperature rather than
energy, can be obtained from Eq.\ (\ref{eq:RRKM}) by Boltzmann weighting thus:
\begin{equation}
k^\dag_j(T)=\frac
{\int_{V^\dag}^\infty k^\dag_j(E)\Omega_j(E)e^{-E/\kb T}dE}
{\int_{V_j}^\infty \Omega_j(E)e^{-E/\kb T}dE},
\end{equation}
where $V_j$ is the potential energy of minimum $j$. Since the Laplace transforms
\begin{equation}
Z_j(T)=\int_{V_j}^\infty\Omega_j(E)e^{-E/\kb T}dE
\end{equation}
and
\begin{equation}
Z^\dag(T)=\int_{V^\dag}^\infty\Omega^\dag(E)e^{-E/\kb T}dE
\end{equation}
are the vibrational partition functions of the minimum and transition state respectively,
we obtain
\begin{equation}
k^\dag_j(T)=\frac{\kb T}{h}\frac{Z^\dag(T)}{Z_j(T)}.
\label{eq:TST}
\end{equation}
\par
Assigning a density of states or partition function to each minimum implies that
the equilibrium properties of the system are described by the superposition
principle\cite{Hoare79a,frankehb93,Wales93f}.
The total density of states and partition function are given by
\begin{equation}
\Omega(E)=\sum_{i=1}^{n_{\rm min}}\Omega_i(E) \quad\text{and}\quad
Z(T)=\sum_{i=1}^{n_{\rm min}}Z_i(T).
\end{equation}
Application of detailed balance to Eqs.\ (\ref{eq:RRKM}) and (\ref{eq:TST}) supplies the equilibrium
occupation probabilities in the microcanonical and canonical ensembles:
\begin{equation}
P_i^{\rm eq}(E)=\Omega_i(E)/\Omega(E), \quad
P_i^{\rm eq}(T)=Z_i(T)/Z(T).
\label{eq:eqmprobs}
\end{equation}
These components emerge from the master equation as $t\to\infty$. 
\par
The use of $\Omega_i(E)$ and $Z_i(T)$ to calculate rates for the master equation when
${\bf P}\ne{\bf P}^{\rm eq}$ [i.e., when Eqs.\ (\ref{eq:eqmprobs}) do not hold] still requires thermal
equilibration of phase points {\em within} each potential well. This condition is fulfilled
under the time scale separation of inter- and intrawell motion mentioned at the end of
Sec.\ \ref{sect:master}.

\section{Morse Clusters}
\label{sect:morse}

In this section, we apply the master equation to databases derived for the 13-atom
Morse cluster, M$_{13}$. The Morse potential\cite{Morse29a} can be written as
\begin{equation}
V=\sum_{i<j}V_{ij};\hspace*{10mm}
V_{ij}=e^{\rho(1-r_{ij}/r_{\rm e})}
[e^{\rho(1-r_{ij}/r_{\rm e})}-2]\epsilon,
\end{equation}
where $r_{\rm e}$ and $\epsilon$ are the dimer pair separation and well depth respectively.
$\rho$ is a dimensionless parameter which determines the range of the interparticle
forces, with small $\rho$ corresponding to long range. Physically meaningful values
include $\rho=3.15$ and $3.17$ for sodium and potassium\cite{Girifalco59a}
up to about $14$ for C$_{60}$ molecules\cite{Girifalco92a,Wales94c}. When $\rho=6$, the
Morse potential has the same curvature as the Lennard-Jones potential at the minimum.
$\epsilon$, $r_{\rm e}$ and the atomic mass, $m$, define the system of reduced units
for the Morse potential independently of $\rho$, giving $\mtime$ as the reduced unit
of time.
\par
Details of databases obtained for four values of $\rho$ have been given
previously\cite{Miller99a}. They demonstrate that the energy landscape of these
clusters is a single funnel---a collection of kinetic pathways leading to a particular
low-energy structure. However, the funnel has a steeper energetic gradient
and lower downhill barriers when the range of the potential is long. The
number of stationary points increases dramatically as the range is decreased, and the
system must then overcome more barriers to reach the global minimum. These effects are
expected to hinder relaxation towards the global minimum. In Sec.\ A, we examine the
performance of some partition function models for the equilibrium properties of the
clusters, before addressing the dynamics in Sec.\ B.

\subsection{Partition Function Models}

The ingredients for the equations of Sec.\ \ref{sect:rateseqm} are the densities of states
for each minimum and transition state. If the harmonic approximation is applied to the
vibrational density of states of a one-component $N$-atom cluster, the result is
\begin{equation}
\Omega^{\rm HO}_i(E)=\frac{2N!}{h^{\rm PG}_i}
\frac{(E-V_i)^{\kappa-1}}{(\kappa-1)!\,(h\bar\nu_i)^\kappa},
\label{eq:HODOS}
\end{equation}
where $\bar\nu$ is the geometric mean frequency, $\kappa=3N-6$ is
the number of vibrational degrees of freedom and $h^{\rm PG}_i$ is the order of the point group
of the isomer. The corresponding partition function is
\begin{equation}
Z^{\rm HO}_i(T)=\frac{2N!}{h^{\rm PG}_i}\frac{(\kb T)^\kappa}{(h\bar\nu_i)^\kappa}e^{-V_j/\kb T}.
\label{eq:HOpartition}
\end{equation}
Eqs.\ (\ref{eq:HODOS}) and (\ref{eq:HOpartition}) can be applied to transition
states by excluding the reactive mode, in which case $\kappa=3N-7$.

The harmonic treatment is approximate in two ways. Firstly, the anharmonicity of the
potential is ignored. Secondly, the density of states around each minimum is
modelled as the surface area of a hyperellipsoid in phase space, and the overlap of
the hyperellipsoids belonging to different minima is neglected. Both these effects
become more pronounced as the total energy increases; highly anharmonic barrier
regions are reached, and the phase space hyperellipsoids become larger.

\end{multicols}
\begin{center}
\begin{figure}
\epsfig{figure=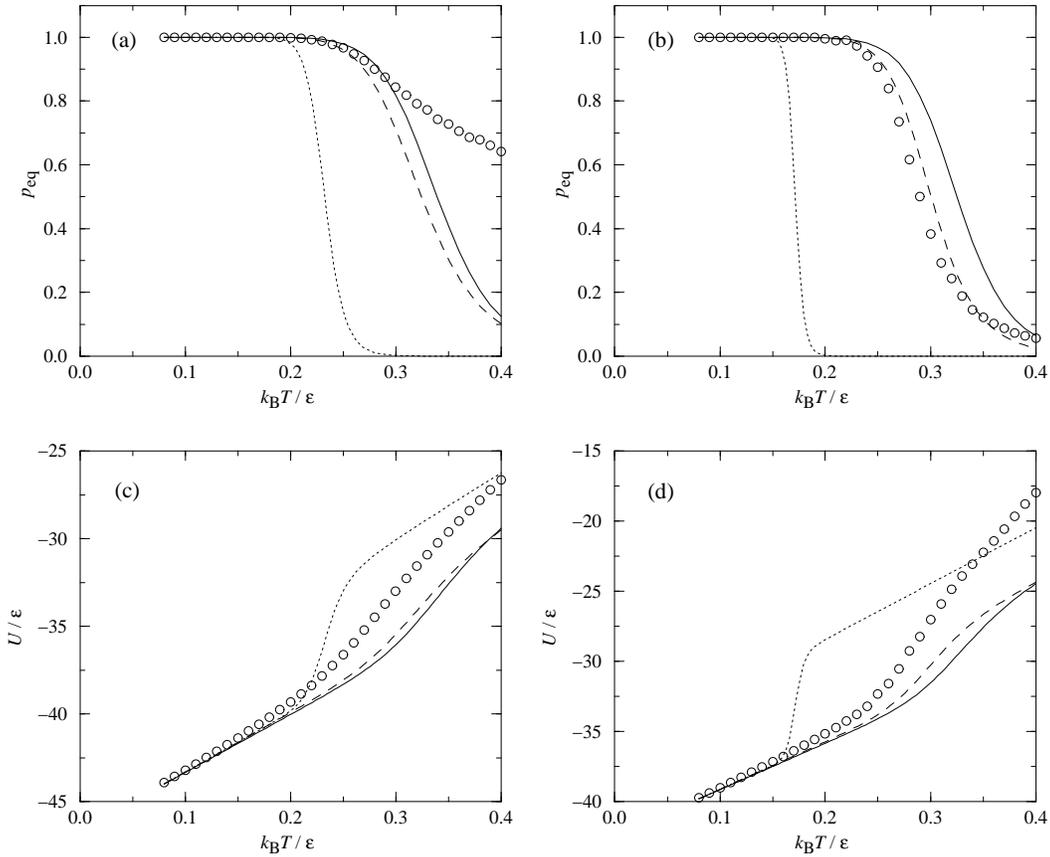,width=140mm}
\begin{minipage}{175mm}
\caption{
Equilibrium occupation probability of the global minimum (top row) and
canonical caloric curves (bottom row) for M$_{13}$ with $\rho=4$ (left column)
and $\rho=6$ (right column). Circles: canonical MC simulations,
solid lines: harmonic approximation, dotted lines: MB method, dashed lines:
MB$(\eta_{\rm P}=0.1)$ method (see text).
}
\label{fig:pftest}
\end{minipage}
\end{figure}
\end{center}
\begin{multicols}{2}

A simple test of the partition function model is to compare the equilibrium
probability of the global minimum predicted by Eqs.\ (\ref{eq:eqmprobs}) with the fraction
of quenches to this structure in the course of a simulation. We have observed elsewhere\cite{Miller97a}
that the harmonic approximation works quite well for the equilibrium probabilities
of LJ$_7$ isomers in the microcanonical ensemble, even in the liquid-like regime.
Figures \ref{fig:pftest}a and \ref{fig:pftest}b show how the harmonic approximation performs for the
global minimum of the M$_{13}$ clusters with $\rho=4$ and 6 in the canonical ensemble, using
the databases obtained previously\cite{Miller99a}. The graphs also show the equilibrium probability
of the global minimum obtained by quenching from canonical Monte Carlo (MC) simulations. In the
simulations, a spherical container of radius $3\,\sigma$ was imposed to prevent evaporation.
At each temperature $2\times10^6$ single-particle warm-up steps were performed before collecting
data over $2\times10^8$ steps, quenching every $10^4$ steps to find the local minimum in which the phase
point resided. The harmonic curves depart from the MC results soon after the cluster begins to melt.
For $\rho=6$ (Fig.\ \ref{fig:pftest}b), the harmonic curve has the correct qualitative shape,
but is shifted to higher temperature. For $\rho=4$ (Fig.\ \ref{fig:pftest}a), however, the harmonic
approximation underestimates the MC result increasingly badly above about $\kb T=0.3\,\epsilon$.

The modelling of partition functions of individual minima for use in the master
equation has been the subject of an extensive study by Ball and Berry\cite{Ball98a,Ball98b}.
These authors considered a variety of analytic forms for $Z_i(T)$ which
attempt to improve on the shortcomings of the harmonic approximation.
The greatest improvement was produced using
an anharmonic model based on a first order expansion of the density of states for the Morse
potential\cite{Haarhoff63a,Doye95a}. The anharmonic correction for each minimum was derived
from the heights of the barriers connected to it. The resulting partition function
for this ``Morse barrier'' (MB) method is
\begin{equation}
Z^{\rm MB}_i(T)=Z^{\rm HO}_i\prod_{j=1}^{n_i}\left(1+a^{(i)}_j\kb T\right)^{\alpha_i}.
\label{eq:MBpf}
\end{equation}
In this expression, $n_i$ is the number of transition states connected to minimum $i$, and
$a^{(i)}_j=(2\Delta V^{(i)}_j)^{-1}$ is the anharmonicity parameter derived from the
barrier height, $\Delta V^{(i)}_j$, of the $j$th transition state connected to minimum
$i$. The power $\alpha_i={\rm min}[1,\kappa/n_i]$ compensates for the possibility
that there may be more than $\kappa$ transition states connected to a given minimum,
since the original model on which Eq.\ (\ref{eq:MBpf}) is based associated one barrier with each
normal mode. The MB model experiences problems when minima are connected by low barriers
because the full series for the density of states of a Morse oscillator, from which the
partition function is derived, diverges as $\kb T$ approaches $\Delta V^{(i)}_j$.
This behaviour correctly corresponds to dissociation of the Morse oscillator, but for
a cluster isomerisation the density of states should remain finite above the barrier.
Ball and Berry\cite{Ball98a} suggested two ways to circumvent this difficulty. In the
first approach, anharmonic corrections were applied only to low-energy minima, which tend
not to have small barriers. This prescription is not entirely satisfactory because the low
barriers of high-energy minima are a signature of the strong anharmonicity in these states,
which is then ignored by this method. Furthermore, some criterion for selecting minima
for anharmonic corrections is required. The second method, called MB$(\eta_{\rm P})$,
involved limiting the anharmonic correction $a^{(i)}_j\kb T$ for each barrier to a plateau
value, $\eta_{\rm P}$, at temperatures where it would exceed this value, i.e.,
\begin{equation}
Z^{{\rm MB}(\eta_{\rm P})}_i(T)=Z^{\rm HO}_i\prod_{j=1}^{n_i}
\left(1+{\rm min}\left[\eta_{\rm P},a^{(i)}_j\kb T\right]\right)^{\alpha_i}.
\label{eq:MBP}
\end{equation}
Although a reasonable value of $\eta_{\rm P}$ might be expected to be around
$0.5$ (i.e., where $\kb T=\Delta V^{(i)}_j$), the best agreement with quenching
from constant temperature molecular dynamics (MD) simulations was achieved for $\eta_{\rm P}\approx 0.1$.
\par
The global minimum equilibrium probabilities of the M$_{13}$
clusters with $\rho=4$ and 6 using the MB and MB$(\eta_{\rm P})$ formulations
are shown by the dotted and dashed lines in
Figs.\ \ref{fig:pftest}a and \ref{fig:pftest}b. The inadequacy of the unconstrained MB model is obvious;
the large anharmonic corrections of the high-energy minima cause these
states to dominate as soon as they become energetically accessible, making the probability
of the global minimum plummet at an artificially low temperature. For $\rho=4$, the
MB$(\eta_{\rm P})$ model produces a marginally better match to the MC data at low temperature
than the harmonic approximation, but then deviates more rapidly. For $\rho=6$, quite a
reasonable improvement is achieved. However, the position of the curve is continuously
adjustable from the MB method (effectively $\eta_{\rm P}=\infty$) to the harmonic approximation
($\eta_{\rm P}=0$), so it is difficult to argue that the improvement is based on physical
insight.
\par
The effect of the partition function model on the equilibrium probabilities of minima
other than the global minimum is hard to gauge for a system where there are so many
of them, since the quench statistics are poor.
However, an impression of the overall description of the PES can be gained through
thermodynamic properties derived from the superposition method. The internal energy
can be obtained from the standard relation $U=\kb T^2(\partial\ln Z/\partial T)_{N,V}$.
The harmonic approximation yields the classical equipartition result
\begin{equation}
U^{\rm HO}=\frac{1}{Z^{\rm HO}}\sum_{i=1}^{n_{\rm min}}Z^{\rm HO}_i\left(V_i+\kappa\kb T\right),
\label{eq:UHO}
\end{equation}
and the MB$(\eta_{\rm P})$ model [Eq.\ (\ref{eq:MBP})] gives
\begin{eqnarray}
U^{{\rm MB}(\eta_{\rm P})}=\frac{1}{Z^{{\rm MB}(\eta_{\rm P})}}
\sum_{i=1}^{n_{\rm min}} Z^{{\rm MB}(\eta_{\rm P})}_i\Bigg[V_i+\kappa\kb T \nonumber \\
+\alpha_i(\kb T)^2\sum_{j=1}^{n_i}\frac{a^{(i)}_j\Theta\left(\eta_{\rm P}-a^{(i)}_j\kb T\right)}
{1+a^{(i)}_j \kb T}\Bigg],
\label{eq:UMBP}
\end{eqnarray}
where the step function signifies that the derivative of the anharmonic correction from a given
mode to the partition function is zero once the plateau has been reached. The derivative is
also assumed to vanish at the point $a^{(i)}_j\kb T=\eta_{\rm P}$, where the plateau
starts, even though it is really discontinuous.
\par
Figures \ref{fig:pftest}c and \ref{fig:pftest}d compare the canonical caloric curves of M$_{13}$ with $\rho=4$
and 6 given by Eqs.\ (\ref{eq:UHO}) and (\ref{eq:UMBP}). Also shown are the results from MC simulation,
obtained by adding the kinetic contribution $\kappa\kb T/2$ to the configurational part
given by the simulations. For both clusters, the harmonic approximation underestimates the
internal energy. The MB method incorrectly predicts a sharp transition at the temperatures
where the equilibrium probability shifts from the global minimum in Figs. \ref{fig:pftest}a
and \ref{fig:pftest}b. The MB$(\eta_{\rm P}=0.1)$ method, however, represents only a marginal improvement
on the harmonic approximation. We have also found this to be the case for the LJ$_9$ cluster
studied by Ball and Berry\cite{Ball98a}, using the model that gave the greatest improvement
for the equilibrium probabilities of the various isomers. This result shows that optimizing the
models for the probabilities does not necessarily produce the correct thermodynamic behaviour.
Doye and Wales found that, whilst first order anharmonic corrections were enough to
improve the thermodynamic description of LJ$_{55}$ in the superposition method, second order
corrections were necessary for the smaller LJ$_{13}$\cite{Doye95a}.
\par
Attempts to model the density of states of individual minima using an analytic function of the
energy implicitly assume that the ``shape'' of the basin of attraction is simple enough to be
described adequately by such a form. In practice, basins of attraction are probably
highly complex objects. For example, in previous work\cite{Miller99a} we saw that increasing the range of the
potential removes locally stable minima, but remnants of these features are likely
to persist as shoulders or inflections on the PES, so that regions of configuration
space that were associated with shallow minima for a shorter-ranged potential become
formally associated with other minima when the range is increased. These features might
explain why the model partition function results in Fig.\ \ref{fig:pftest} are worse for $\rho=4$ than
for $\rho=6$.
\par
\begin{figure}
\begin{center}
\epsfig{figure=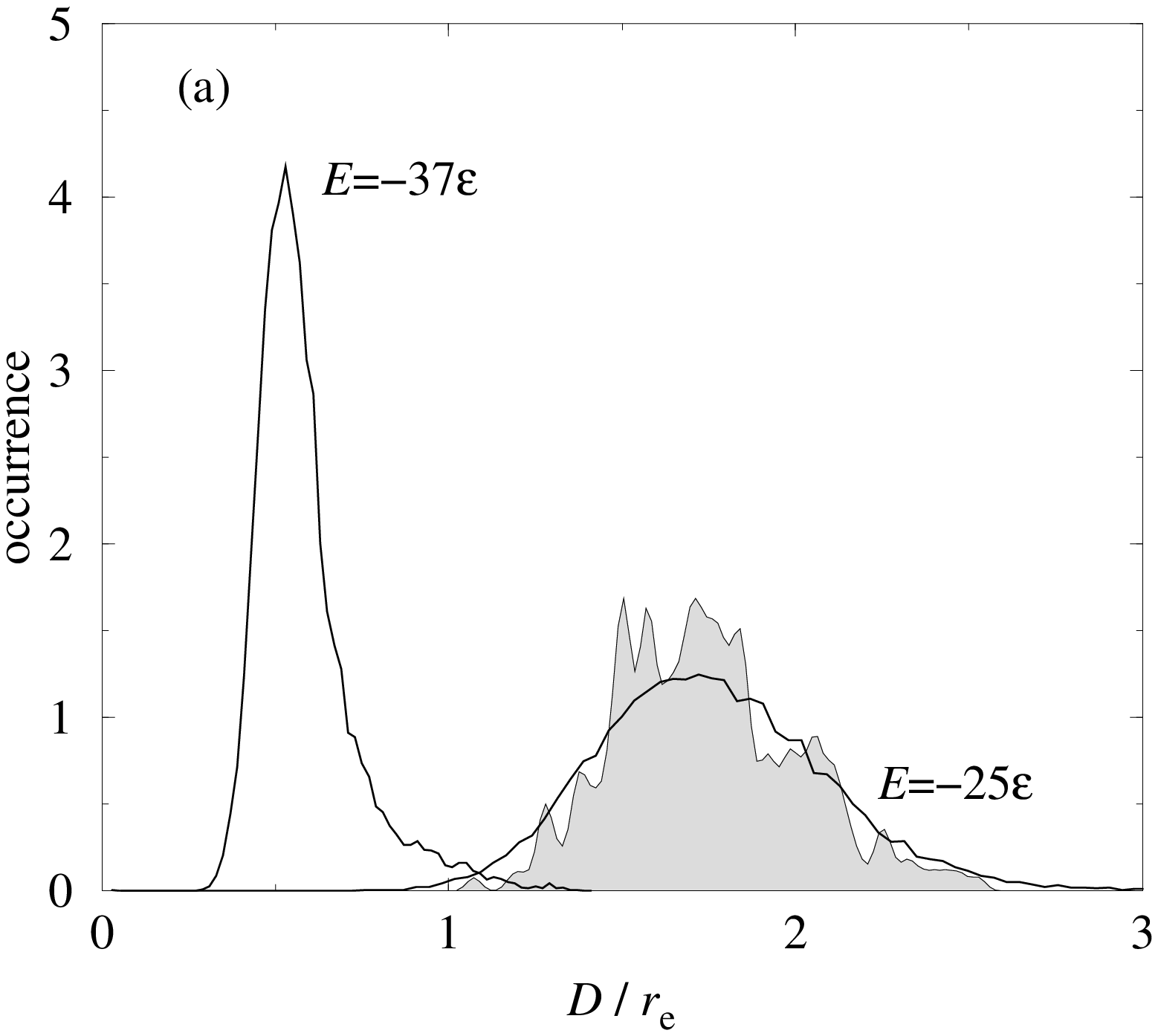,width=76mm} \\
\epsfig{figure=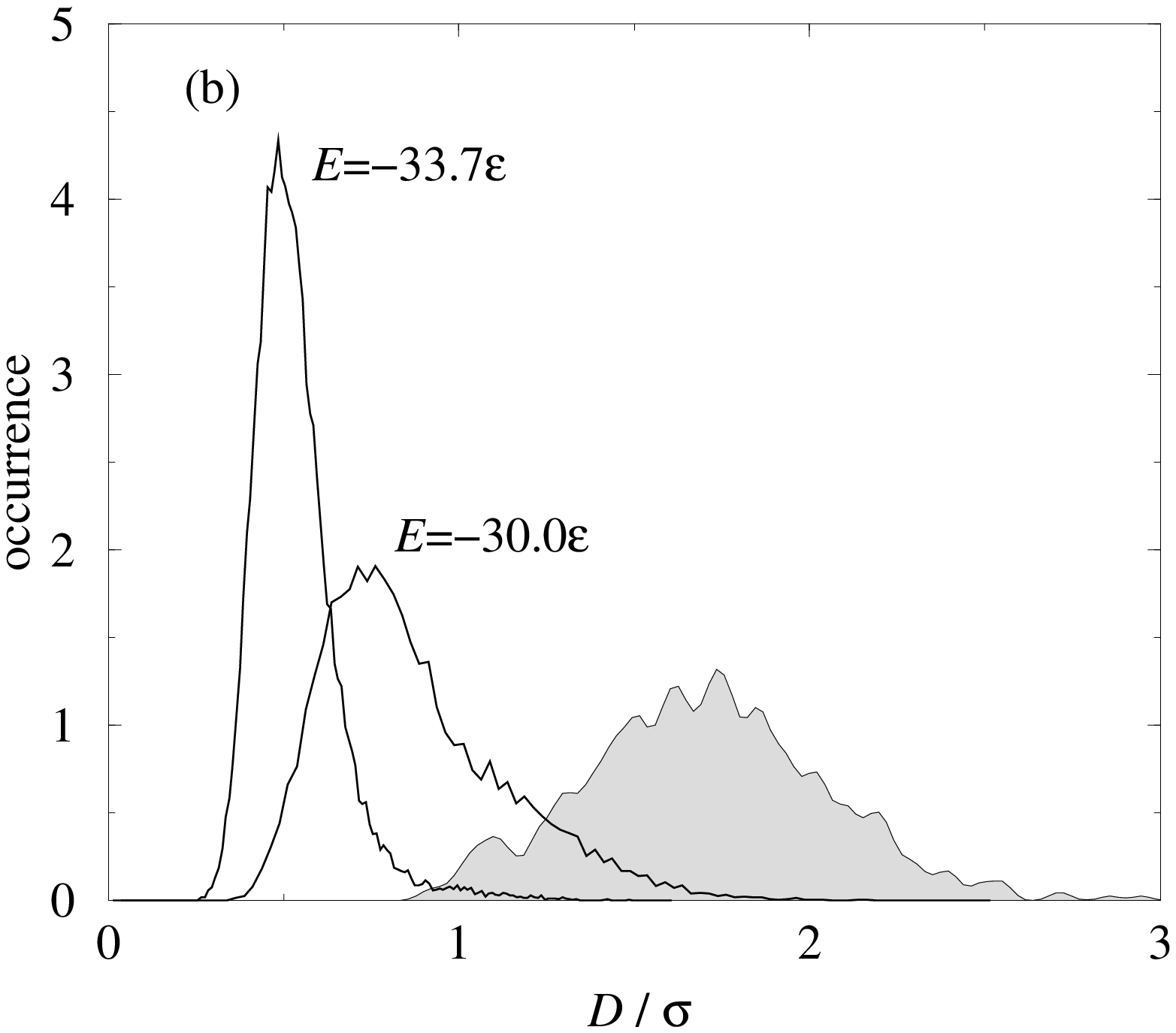,width=76mm}
\end{center}
\begin{minipage}{85mm}
\caption{
Distribution of Euclidean distances to the global minimum from configuration points
in its basin of attraction for (a) M$_{13}$, $\rho=4$ and (b) LJ$_{13}$. For each
cluster, the distributions from microcanonical MD simulations at two energies are
shown. At the lower energy, the majority of quenches lead to the global minimum,
and at the higher energy, about half do so. The quench interval was 15 reduced time units.
The duration of the high-energy simulations was $6\times10^5$, and that of the
low-energy ones was $3\times10^5$. The shaded areas show the distance distribution
of minima that are directly connected to the global minimum.
}
\label{fig:quenchdist}
\end{minipage}
\end{figure}

To illustrate this effect, we have performed microcanonical MD simulations of
M$_{13}$ with $\rho=4$, and LJ$_{13}$, which closely resembles M$_{13}$ with $\rho=6$.
Periodic quenching was applied, and the Euclidean distance, $D$, was calculated between the
configuration point taken at the start of the quench and the local minimum to which it converged.
The distribution of $D$ for the subset of quenches that led to the global minimum is plotted
in Fig.\ \ref{fig:quenchdist} at two energies for each cluster.
In each case, the lower energy has been chosen just
above that at which the trajectory can escape from the global minimum, so that over
95\% of quenches return to the global minimum, and the cluster is exploring a large proportion
of the catchment basin of this structure. The higher energy was chosen such that about
half the quenches return to the global minimum. The corresponding distributions are broader
and peak at a higher value of $D$, as would be expected. The shaded region on each graph
shows the distribution of $D$ for the minima directly connected to the global minimum. Each
of these minima is surrounded by its own basin of attraction, and a barrier must be surmounted
before the configuration point enters the catchment basin of the global minimum. Even so,
in Fig.\ \ref{fig:quenchdist}b for LJ$_{13}$, the tail of the high-energy quench distribution
overlaps somewhat with the distribution of connected minima, indicating that some of the points in
the basin of attraction of the global minimum are as far from it as the closest local minima.
In Fig.\ \ref{fig:quenchdist}a for M$_{13}$ with $\rho=4$, however, the high-energy quench distribution
overlaps completely with the distribution of connected minima. The overlap means that many
configurations which differ structurally from the global minimum as much as the connected local minima
still quench to the global minimum. Such configurations may include structures
which for a slightly shorter-ranged potential fall into the catchment basin of a different
local minimum.

Further evidence to support this explanation comes from Fig.\ \ref{fig:pftest}. The MC results for
$\rho=6$ show that the increase in gradient of the caloric curve---indicative of the system
sampling a new region of configuration space---occurs at the temperature where the trajectory
begins to escape from the global minimum, as shown by the decrease in $P^{\rm eq}_{\rm gmin}$.
The transition is therefore out of the basin of attraction of the global minimum, effectively
from solid-like to liquid-like states. The analogous results for $\rho=4$ show that the
transition feature in the caloric curve starts before the probability of the global minimum
drops significantly. This behaviour is suggestive of a weak transition from configurations close
to the global minimum to higher-energy ones, all taking place within the basin of attraction of
the global minimum itself.
\par
The picture of the basin of attraction around the global minimum that emerges
is therefore complex. It extends far into configuration space from the icosahedron,
and past the catchment basins of local minima in certain directions. The structural
dissimilarity of some points which are formally associated with the global minimum suggests
that quenching might not be the most meaningful way of dividing configuration space amongst
the various minima. From this point of view, the underestimation of the global minimum
probability in Fig.\ \ref{fig:pftest}a by the harmonic superposition method is somewhat misleading,
since it is not helpful to think of the distant configurations as belonging to the
icosahedral well. The harmonic approximation may therefore give a more meaningful
probability that the cluster has a structure resembling the global minimum than quenching.

\end{multicols}
\begin{center}
\begin{figure}
\epsfig{figure=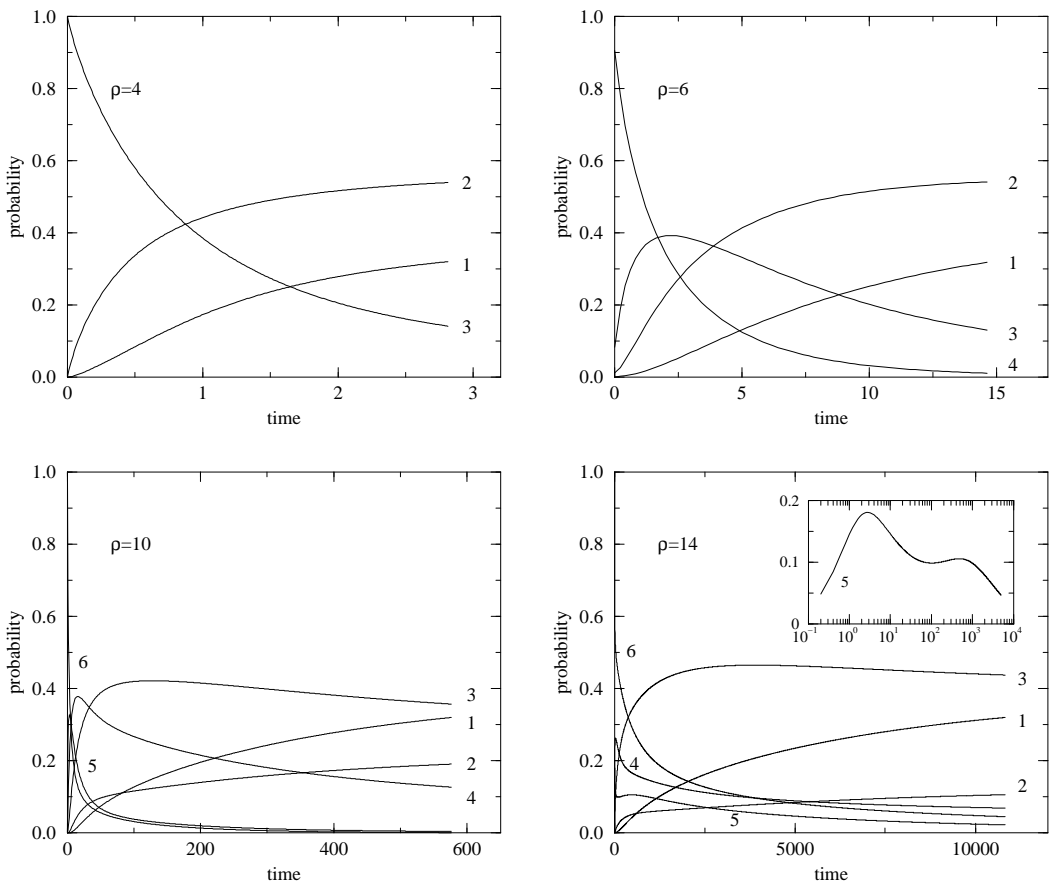,width=140mm}
\begin{minipage}{175mm}
\caption{
Relaxation of minima, grouped in ``layers'' away from the global minimum, for M$_{13}$ at four values
of the range parameter, $\rho$. In each plot, layer 1 is the global minimum, layer
2 contains all minima directly connected to layer 1, etc. In each case, the microcanonical
total energy is chosen such that the equilibrium probability of the global minimum
is $0.4$: for $\rho=4,6,10,14$, $E/\epsilon=-33.17,-29.42,-28.43,-29.78$, respectively. The inset
for $\rho=14$ shows the first half of the layer 5 curve with a logarithmic time axis.
The time is in units of $\mtime$.
}
\label{fig:layers}
\end{minipage}
\end{figure}
\end{center}
\begin{multicols}{2}

In summary, although the harmonic approximation has only partial success in describing the
equilibrium properties of the clusters examined here, it is attractive in its simplicity,
lack of empirical parameters, and clear physical basis. More complicated analytic
models do not necessarily provide greater insight, or even systematically improved results.
We therefore adopt the harmonic approximation for the rate constant and equilibrium
property expressions in Sec.\ \ref{sect:rateseqm}, with the proviso that they will only be applied
at low and moderate temperatures, where the description should be adequate for our purposes.
The resulting microcanonical rate constants, via Eq.\ (\ref{eq:RRKM}), are given by 
\begin{equation}
k^\dag_i(E)=\frac{h^{\rm PG}_i}{h^{\rm PG\dag}}
\frac{\bar\nu_i^\kappa}{\bar\nu^{\dag(\kappa-1)}}
\left(\frac{E-V^\dag}{E-V_i}\right)^{\kappa-1},
\label{eq:RRK}
\end{equation}
and the canonical expression from Eq.\ (\ref{eq:TST}) is
\begin{equation}
k^\dag_j(T)=\frac{h^{\rm PG}_j}{h^{\rm PG\dag}}
\frac{\bar\nu_i^\kappa}{\bar\nu^{\dag(\kappa-1)}}
e^{-(V^{\dag}-V_j)/\kb T}.
\label{eq:HOTST}
\end{equation}
Finally, we note that the harmonic superposition method is likely to be more successful in
the canonical ensemble than the microcanonical because at constant temperature
the velocity distribution is independent of the configuration point. In contrast, at
fixed total energy the kinetic energy is significantly further ``above'' the PES when the
configuration point is near a deep potential well than when it is near higher-lying ones.

\subsection{Relaxation and the Range of the Potential\label{sect:rangerelax}}

\subsubsection{Relaxation to the Global Minimum}

We now turn to the effect of the range of the potential on the dynamics of the M$_{13}$
clusters in the light of our previous analysis of the energy landscapes\cite{Miller99a}.
The larger number of rearrangements and smaller energy gradient on paths to
the global minimum, as well as the higher downhill barriers, are expected to
impede relaxation to the global minimum as the range of the potential is decreased.
\par
Figure \ref{fig:layers} illustrates the range dependence of structural relaxation to the global
minimum. The minima are grouped into ``layers''
according to the smallest number of rearrangements required to reach the global minimum.
Layer 1 contains only the global minimum, and level $i+1$ contains all minima directly
connected to a minimum in level $i$, but not to any minimum in a layer lower than $i$.
The initial probability vector at each value
of $\rho$ is a uniform distribution amongst the minima in the furthest layer from the
global minimum. The energy has been chosen such that the equilibrium probability of the
global minimum is $0.4$, and the solution of the master equation is shown until the time at
which 80\% of this population (i.e., $0.32$) has been achieved.
\par
The most striking trend is the increasing time scale as the potential range decreases,
as predicted by the landscape analysis\cite{Miller99a}. The $\rho=14$ cluster takes over
three orders of magnitude longer to reach 80\% of equilibrium than the $\rho=4$ cluster.
Note that for the latter system, the time scale plotted is only of the order of a few vibrational
periods\cite{Miller99a}. In the light of previous work\cite{Miller97a}
the application of the master equation to the $\rho=4$ cluster is therefore
probably at the limit of validity, since the assumption of stochastic transitions
breaks down when the relaxation time approaches the vibrational period.
\par
The relaxation is straightforward for $\rho=4$; the furthest layer decays whilst the
populations of the global minimum and the intermediate layer grow monotonically.
At $\rho=6$ we see the accumulation and decay of a transient population in
layer 3 as the outward flow from these minima does not match the rapid initial input
from the furthest layer. The layer 3 minima therefore constitute a kinetic bottleneck
for relaxation down the funnel of the PES. More complex behaviour arises for
$\rho=10$ and 14, where there are six layers of minima. The probabilities experience
an initial jump as the system is released from its strongly non-equilibrium
state, and then relax slowly to their final values. Layer 5 at $\rho=14$ (shown in the
inset of Fig.\ \ref{fig:layers}) shows particularly
complicated behaviour, rising suddenly at first, decaying slightly, and then rising
again before relaxing monotonically. This oscillation occurs because layer 4 develops a transient
population, blocking further downward output from layer 5 while layer 6 is still releasing
probability into layer 5 from above.

\subsubsection{Search Times}

By analogy with the ``folding time'' in the protein folding literature, we can examine
the ability of the cluster to find its global minimum by defining a ``search time'' as the
time taken for the probability of the global minimum to reach a particular value after the
system is released from a non-equilibrium state. Here we use a probability threshold of $0.4$,
but we will discuss the effects of changing this choice.
\par
\begin{center}
\begin{figure}
\epsfig{figure=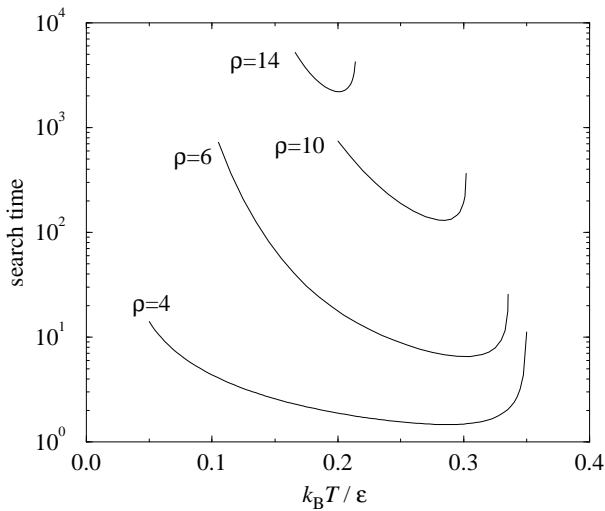,width=80mm}
\begin{minipage}{85mm}
\caption{
Search time as a function of temperature for M$_{13}$ at four values of the
range parameter, $\rho$. The search time is defined as the time taken for the
probability of the global minimum to reach $0.4$, starting from an even probability
distribution amongst the minima in the layer furthest from the global minimum.
The time is in units of $\mtime$.
}
\label{fig:foldingtime}
\end{minipage}
\end{figure}
\end{center}

Fig.\ \ref{fig:foldingtime} shows the search time for M$_{13}$ as a function of temperature
using four values of $\rho$. The initial probability vector was a uniform distribution
amongst the minima in the layer furthest from the global minimum. The qualitative shape
of the curves is easily understood. As the temperature is increased from low values,
the search time decreases because the thermal energy rises above the barriers between
the minima. An optimal temperature, $T_{\rm opt}$ is reached, where the search time is a
minimum, $\tau_{\rm opt}$, above which it rises because the thermodynamic
driving force towards the global minimum is reduced at higher temperatures. Ultimately,
the equilibrium probability of the global minimum falls below the threshold of $0.4$,
and the search time is no longer defined. Similar behaviour has been observed for the
same reasons in direct simulations of KCl clusters\cite{Rose93b} and lattice protein
models\cite{Socci96a}, as well as a master equation study of idealized energy
landscapes\cite{Doye96b}.

Table \ref{tab:fold} lists the temperature and $\tau_{\rm opt}$, as well as the
two temperatures, $T_{\rm low}$ and $T_{\rm high}$, at which the search time equals
$2\tau_{\rm opt}$. The difference $T_{\rm high}-T_{\rm low}$ provides a measure of
the width of the temperature window for which searching is reasonably fast.
The value of $\tau_{\rm opt}$ increases with $\rho$, as expected from the relaxation
profiles of the previous section, and this slowing down is accompanied by a decrease
in the temperature width $T_{\rm high}-T_{\rm low}$. As the range of the potential
is decreased, the energy gap between the global minimum and the other minima becomes
smaller, and the energy range spanned by the minima narrows. Hence, other minima
come into play at lower temperature, and the temperature at which the global minimum
ceases to dominate the equilibrium populations is lower when the potential is short-ranged.
This observation explains why $T_{\rm high}$ falls as $\rho$ increases. For high $\rho$, the
downhill barriers between minima are on average larger, so that as the temperature is decreased,
isomerisation processes slow down more dramatically than for small $\rho$. Hence, the search time
increases more rapidly as the temperature is lowered beyond $T_{\rm opt}$ when the potential
is short-ranged, resulting in the narrower ranges of $T_{\rm high}-T_{\rm low}$.

\end{multicols}
\begin{center}
\begin{table}
\begin{minipage}{175mm}
\caption{
``Searching'' characteristics of M$_{13}$ at four values of the range parameter, $\rho$.
$T_{\rm opt}$ is the optimal searching temperature, at which the search time is
a minimum, $\tau_{\rm opt}$. $T_{\rm low}$ and $T_{\rm high}$ ($T_{\rm low}<T_{\rm high}$)
are the two temperatures at which the search time equals $2\tau_{\rm opt}$.
}
\label{tab:fold}
\end{minipage}
\vglue 3mm
\begin{minipage}{110mm}
\begin{tabular}{cccccc}
$\rho$ & $\kb T_{\rm opt}/\epsilon$ &
\multicolumn{1}{c}{$\tau_{\rm opt}\,/{\,(mr_{\rm e}^2/\epsilon)^{1/2}}$} &
$\kb T_{\rm low}/\epsilon$ & $\kb T_{\rm high}/\epsilon$ &
${\displaystyle\frac{T_{\rm opt}-T_{\rm low}}{T_{\rm high}-T_{\rm low}}}$ \\
\hline
4 & 0.293 & 1.47 & 0.135 & 0.344 & 0.76 \\
6 & 0.305 & 6.54 & 0.218 & 0.333 & 0.76 \\
10 & 0.285 & 130 & 0.235 & 0.301 & 0.76 \\
14 & 0.200 & 2200 & 0.170 & 0.214 & 0.68 \\
\end{tabular}
\end{minipage}
\end{table}
\end{center}
\begin{multicols}{2}

An analogy can be drawn here with the ease of folding in proteins. 
The ratio, $T_{\rm f}/T_{\rm g}$, of the ``folding temperature''
(where the native state becomes thermodynamically most stable) to the ``glass transition
temperature'' (where the kinetics slow down dramatically) has been used as a measure
of the ability of a protein to fold correctly\cite{Socci96a,Socci94a}. $T_{\rm f}$ is
roughly related to $T_{\rm high}$, which decreases with increasing $\rho$, whilst $T_{\rm g}$
increases because of the higher barriers for short-ranged potentials. Hence
$T_{\rm f}/T_{\rm g}$ falls, in accordance with the observation that searching for the
global minimum is harder when $\rho$ is high.
\par
Interestingly, the curves in Fig.\ \ref{fig:foldingtime} for $\rho=4$, 6 and 10 have similar shapes.
For example, the values of $T_{\rm opt}$ differ by only 7\%. Furthermore, $T_{\rm opt}$
lies about three quarters of the way from $T_{\rm low}$ to $T_{\rm high}$ in all three cases,
as shown by the last column of Table \ref{tab:fold}. $\rho=14$ represents an extreme case in which relaxation
to the global minimum becomes very slow for values outside a small range near $T_{\rm opt}$. This
optimum temperature is a compromise between the slow dynamics at even moderately low
temperatures, and the rapidly decreasing thermodynamic weight of the global minimum at moderately
high temperatures.
\par
Choosing a different occupation probability of the global minimum as the criterion for the search time has
predictable effects. If a higher threshold is used, the search time increases at any given temperature.
The equilibrium probability of the global minimum drops below the threshold at a lower
temperature, so the upper limit for which the search time is defined decreases. The search time rises
more steeply below $T_{\rm opt}$ because the probabilities must come closer to their equilibrium values,
and this approach is asymptotic. The combined effect is that the search time curves all become narrower.
However, for the cases tested ($\rho=4$ and 6), $T_{\rm opt}$ changed by only 10\% as the threshold was
varied from $0.2$ to $0.5$.

\subsubsection{Relaxation of the Total Energy}

The evolution of the probability vector towards ${\bf P}^{\rm eq}$ is expressed macroscopically
by the relaxation of some overall property, $A$, to its equilibrium value, $A^{\rm eq}$. If this
property has a well defined value, $A_i$, for each state $i$ in the master equation, the expectation
value is a weighted average which can be expressed as a function of time using Eq.\ (\ref{eq:msolution}):
\begin{equation}
\langle A(t)\rangle=\sum_{i=1}^{n_{\rm min}}A_i P_i(t)
=\sum_{j=1}^{n_{\rm min}}c_j e^{\lambda_j t}
=A^{\rm eq}+\sum_{j=2}^{n_{\rm min}}c_j e^{\lambda_j t},
\label{eq:proprelax}
\end{equation}
where
\begin{equation}
c_j=\left[\sum_{i=1}^{n_{\rm min}}A_i\sqrt{P^{\rm eq}_i}\tilde u^{(j)}_i\right]
\left[\sum_{m=1}^{n_{\rm min}}\tilde u^{(j)}_m \frac{P_m(0)}{\sqrt{P^{\rm eq}_m}}\right],
\label{eq:cj}
\end{equation}
and the last expression in Eq.\ (\ref{eq:proprelax}) uses the fact that the $j=1$ term
defines the baseline for relaxation, since $\lambda_1=0$.
A mean relaxation time, $\tau_{\rm r}$, can be defined by normalizing the profile of $\langle A\rangle$
against $t$ (such that it decays from 1 to 0), and evaluating the area under the resulting
curve\cite{Stillinger95a}. For pure Debye (single exponential) relaxation, i.e., $\exp(-\lambda t)$,
one simply obtains $\tau_{\rm r}=\lambda^{-1}$. Subtracting the equilibrium value from the
right-hand side of Eq.\ (\ref{eq:proprelax}), integrating from $t=0$ to $\infty$, and normalizing using the
value at $t=0$ yields
\begin{equation}
\tau_{\rm r}=-\sum_{j=2}^{n_{\rm min}}c_j\lambda_j^{-1}\Big/\sum_{j=2}^{n_{\rm min}}c_j.
\end{equation}
If the eigenvectors and eigenvalues of the transition matrix are not available, $\tau_{\rm r}$
can be obtained by propagating the master equation numerically until $\langle A\rangle$ has
effectively equilibrated, and then numerically integrating the normalized relaxation profile.
Here we will examine the relaxation of the total energy of M$_{13}$ as the populations of the
minima equilibrate at constant temperature. Within the harmonic approximation, therefore, we
need $A_i=V_i+\kappa\kb T$ in Eqs.\ (\ref{eq:proprelax}) and (\ref{eq:cj}).
\par
Theoretical treatments have shown that under particular circumstances, the multi-exponential decay
that arises from the master equation can lead to a variety of asymptotic behaviours. For example,
the random energy model, where the states have a Gaussian distribution of energies, can
lead either to stretched exponential $(\propto\exp[-(t/\tau)^\theta])$ or to power law
$(\propto[\tau/t]^p)$ relaxation of autocorrelation functions,
depending on the form chosen for the transition rates\cite{Koper87a}. Palmer et al.~also
derived stretched exponential behaviour for a hierarchically constrained model\cite{Palmer84a}.
In contrast Skorobogatiy et al.~found power law and logarithmic $(\propto-\ln t)$ decay, but
not stretched exponential decay, of the total energy in a protein model\cite{Skorobogatiy98a},
depending on the temperature.
\par
Starting from a uniform distribution amongst the minima in the layer furthest from the global
minimum, we found that none of the above forms (pure or stretched exponential, power law,
or logarithmic) gave a robust fit to the decay of the total energy from the master equation.
At sufficiently long times, Eq.\ (\ref{eq:proprelax}) approaches pure exponential behaviour, since all
contributions except that of the slowest mode have decayed. At intermediate times, the
parameters (particularly the stretching exponent, $\theta$) that produced the best fit for the
stretched exponential form were highly sensitive to the time interval over which data were supplied,
and the resulting curves often deviated significantly from the master equation solution. The
difficulty of obtaining an acceptable fit increased with decreasing temperature, where the spread
of the exponents $\lambda_j$ is wider. Of course, there is no reason why a general multi-exponential
form like Eq.\ (\ref{eq:proprelax}) should conform to any simplified model.

\begin{figure}
\begin{center}
\epsfig{figure=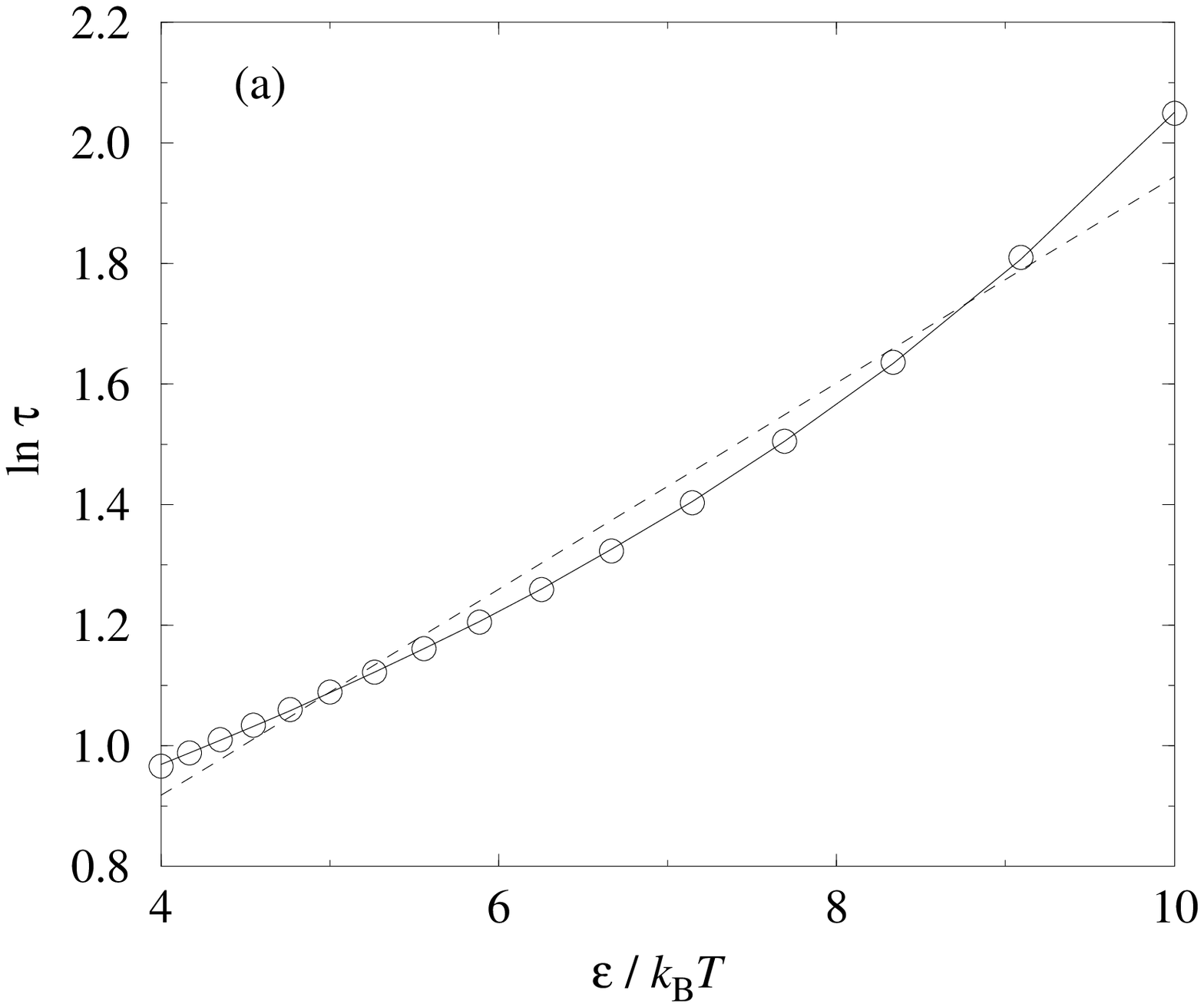,width=76mm} \\
\epsfig{figure=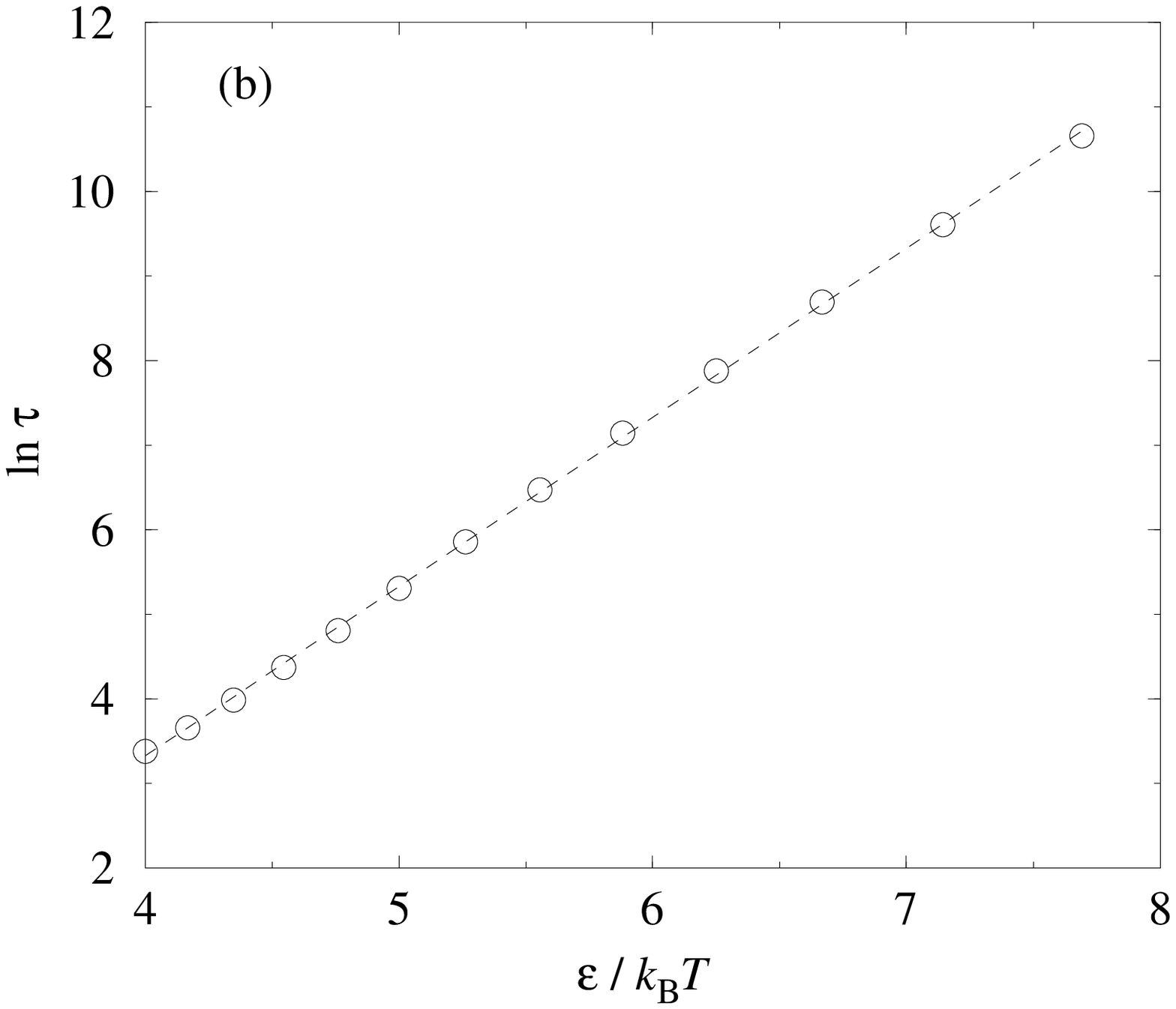,width=76mm}
\end{center}
\begin{minipage}{85mm}
\caption{
Canonical ensemble Arrhenius plots for the relaxation time of the total energy of M$_{13}$
at (a) $\rho=4$ and (b) $\rho=14$. Circles are mean relaxation times from the master
equation, dashed lines are fits to the Arrhenius form, and the solid line in
(a) is a fit to the Vogel--Tammann--Fulcher (VTF) form.
The relaxation time is in units of $\mtime$.
}
\end{minipage}
\label{fig:M13tau}
\end{figure}

Although the relaxation profiles are complicated, the mean relaxation times (integrated
profiles) were found to follow simple empirical expressions. Fig.\ \ref{fig:M13tau}
shows the logarithm of the relaxation
time as a function of inverse temperature for M$_{13}$ with $\rho=4$ and $14$. The $\rho=14$ plot in
Fig.\ \ref{fig:M13tau}b is well fitted by the Arrhenius form $\tau_{\rm r}=\tau_0\exp(A/\kb T)$ with
$\tau_0=9.34\times10^{-3}\mtime$
and $A=2.00\,\epsilon$. The $\rho=4$ plot in Fig.\ \ref{fig:M13tau}a, however,
shows significant deviation from the linearized
Arrhenius expression. It is better fitted by the ubiquitous Vogel--Tammann--Fulcher
(VTF) form\cite{Scherer92a}
\begin{equation}
\tau_{\rm r}=\tau_0\exp\left[\frac{A}{\kb(T-T_0)}\right],
\label{eq:VTF}
\end{equation}
as shown by the solid line, for which the parameters are $\tau_0=1.86\mtime$, 
$A=0.0699\,\epsilon$,
$\kb T_0=0.051\,\epsilon$. These values were found by least squares fitting of the logarithm of
Eq.\ (\ref{eq:VTF}) with equally weighted points.
\par
The slower relaxation and larger database for the $\rho=14$ cluster
meant that it was not computationally feasible to extend Fig.\ \ref{fig:M13tau}b to lower
temperature---the lowest point shown is at $\kb T=0.13\,\epsilon$---and it is possible that deviation from
the Arrhenius behaviour would occur below this value. However, the curvature of the $\rho=4$
plot is clear over the same range, indicating that the relaxation dynamics of the two clusters
respond differently to temperature changes. The origins of the difference probably stem from
the decreasing slope of the energy landscape as the range of the potential is shortened.
The energy intervals spanned by the minimum and transition state samples at higher $\rho$ are narrower,
making the landscape more uniform. This uniformity means that a change in the temperature has
a similar effect on most of the individual interwell processes, each of which separately has an
Arrhenius temperature dependence in the model we have used [Eq.\ (\ref{eq:HOTST})]. On the steeper
landscape of the $\rho=4$ cluster, however, there is a greater spread of local minimum energies and
barrier heights, so that as the temperature is lowered, some processes become ``frozen out''
before others, resulting in longer relaxation times than expected by extrapolation of the
high-temperature behaviour.
\par
In structural glasses, Arrhenius temperature dependence of relaxation times is associated
with strong liquids, whereas VTF behaviour is indicative of fragility \cite{Angell91a}. If this
classification can be applied to clusters, the results of this section suggest that increasing the range
of the potential introduces a degree of fragility. Stillinger's picture of strong liquids having a
``uniformly rough'' energy landscape\cite{Stillinger95a} is in line with our previous analysis\cite{Miller99a},
which showed that decreasing the range of the potential lowers the overall
slope of the PES, leaving a flat but rough landscape.

\subsubsection{Relaxation Modes}

Is the time scale of relaxation mostly determined by the slowest relaxation mode of the
master equation, i.e., the least negative non-zero eigenvalue of the transition matrix,
or is the process it describes relatively unimportant?
Eqs.\ (\ref{eq:proprelax}) and (\ref{eq:cj}) show that the contribution of a given mode
to the relaxation of a global property depends both on the
property and on the initial probability distribution. However, we can still
probe the nature of the probability flow described by a particular relaxation
mode by comparing the size and sign of the components in the corresponding transition
matrix eigenvector. Consider Eq.\ (\ref{eq:msolution}) for a particular value of $j$; mode $j$
makes an important contribution to the probability evolution of minimum $i$ if $u^{(j)}_i$---or
equivalently $\sqrt{P_i^{\rm eq}}\tilde u^{(j)}_i$---is large in
magnitude. The mode represents an overall flow {\em between} minima $i$ and $k$ if
$u^{(j)}_i$ and $u^{(j)}_k$ have opposite signs.
\par
The extreme eigenvalues and eigenvectors of the transition matrix can be obtained
efficiently for large matrices using Lanczos iteration\cite{Golub89a}.
Inspection of the components of the eigenvectors for M$_{13}$ at different values
of $\rho$ and different temperatures reveals some general trends. The extreme
modes (i.e., the slowest and fastest) describe probability flow between a small
number of minima, typically fewer than five. The fastest modes tend to be between
minima that are directly connected by transition states. In contrast, the slowest modes
are between unconnected minima, and probability flow involves intermediate
minima. The slow modes tend to involve one highly populated minimum. Hence, if
the initial probabilities of the other minima that feature in these relaxation modes
(with eigenvector components of opposite sign) are far from their equilibrium values,
the slow modes may limit the overall relaxation.
\par
\begin{figure}
\begin{center}
\epsfig{figure=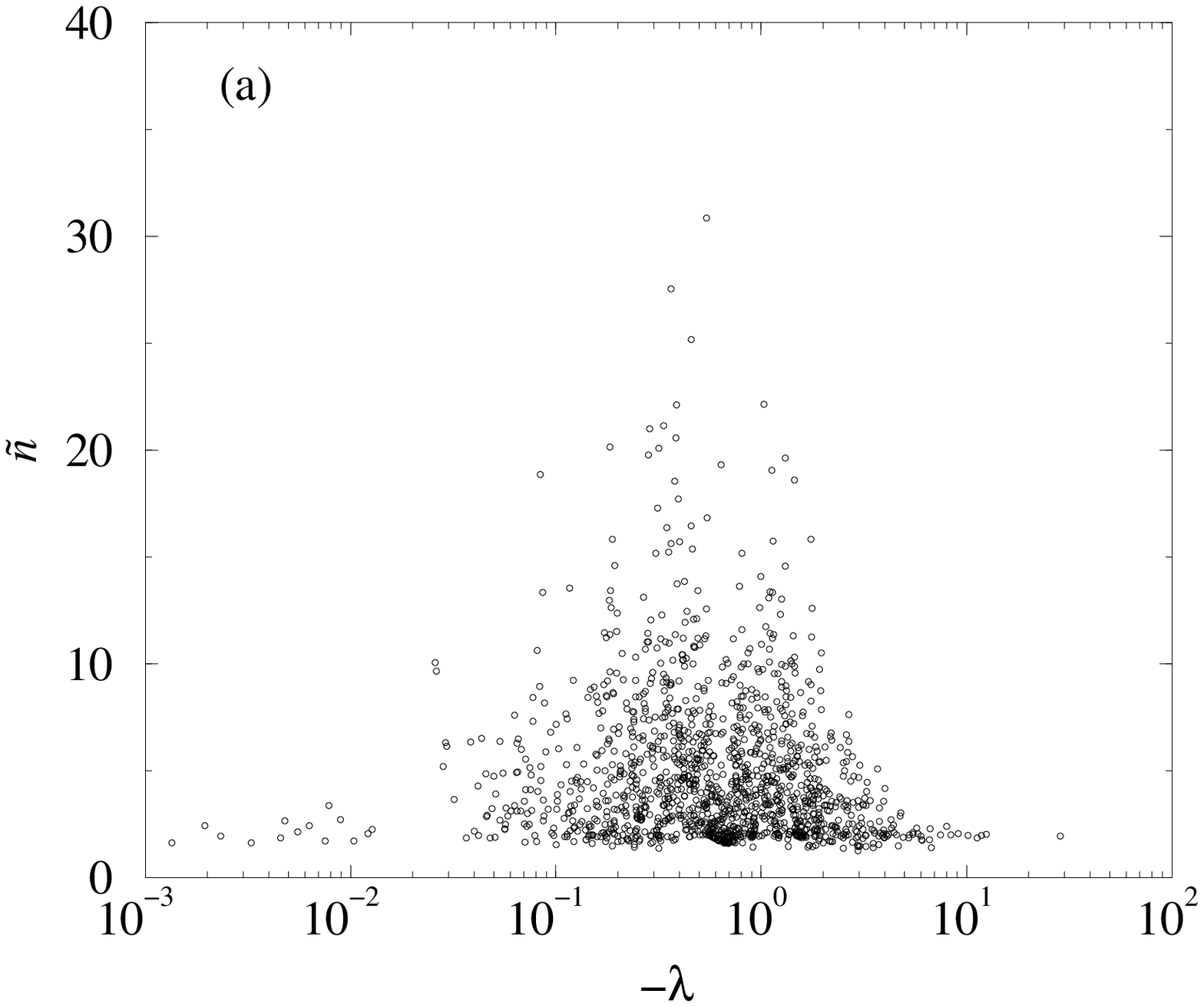,width=76mm} \\
\epsfig{figure=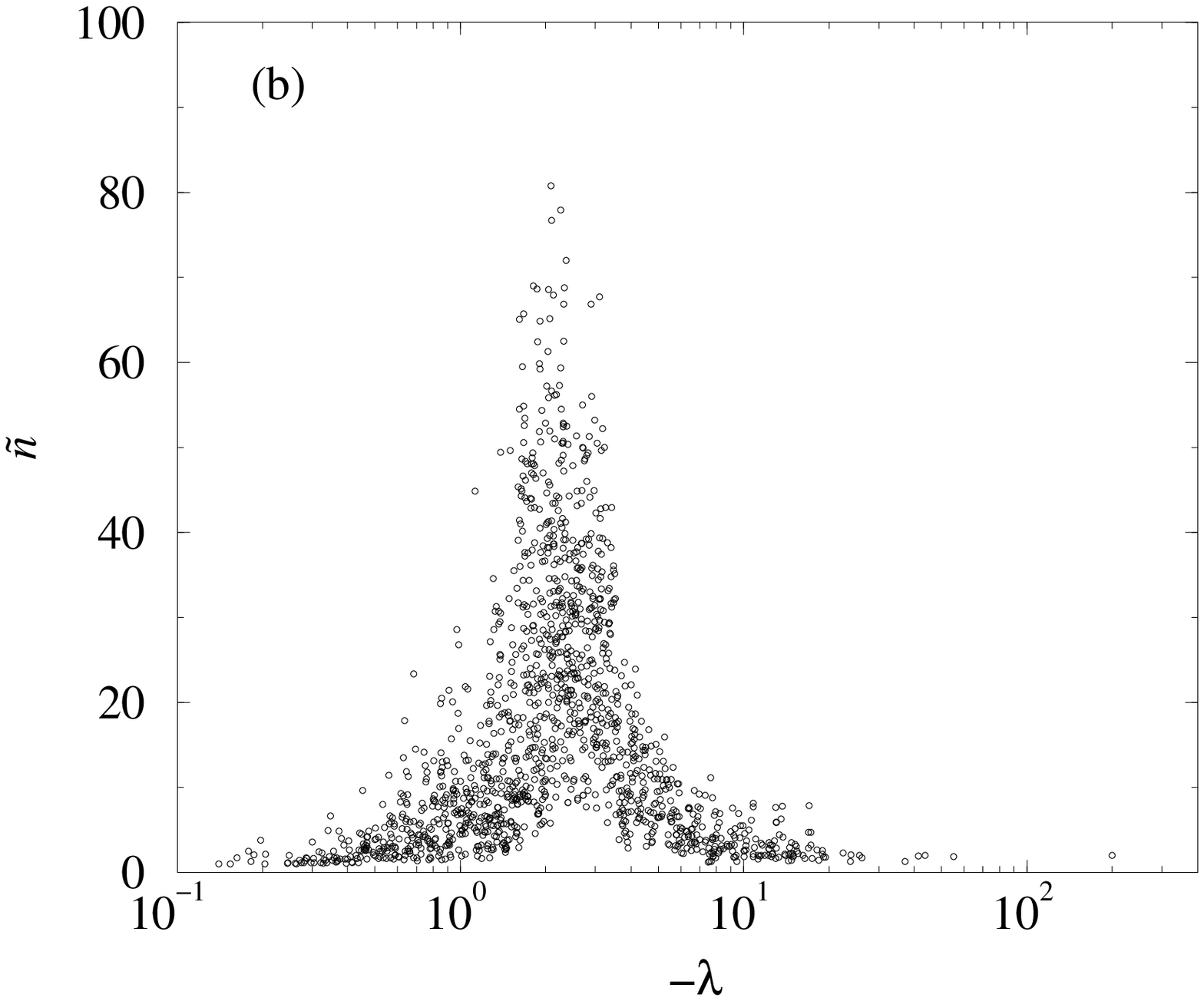,width=76mm}
\end{center}
\begin{minipage}{85mm}
\caption{
Number of minima participating in a relaxation mode of the master equation as a function
of the eigenvalue of the mode for M$_{13}$, $\rho=6$, at
(a) $\kb T=0.15\,\epsilon$ and (b) $\kb T=0.40\,\epsilon$.
The mode with zero eigenvalue has been excluded. $\lambda$ is in
units of $\mfrq$.
}
\label{fig:collect}
\end{minipage}
\end{figure}

The number of minima participating in relaxation mode $j$ can be measured using the
index
\begin{equation}
\tilde n_j=\frac
{\left(\sum_{i=1}^{n_{\rm min}}\left[\tilde u^{(j)}_i\sqrt{P^{\rm eq}_i}\right]^2\right)^2}
{\sum_{i=1}^{n_{\rm min}}\left[\tilde u^{(j)}_i\sqrt{P^{\rm eq}_i}\right]^4},
\end{equation}
which varies from 1 to $n_{\rm min}$. Figure \ref{fig:collect} shows $\tilde n_j$ as a function of the
eigenvalue, $\lambda_j$, for all the relaxation modes of M$_{13}$ with $\rho=6$ at two temperatures.
As observed above, the number of minima involved in the fastest and slowest modes is small.
Many intermediate modes only involve a small number of minima too, but the modes that describe
more global flow are all clustered in the centre of the (logarithmic) scale. This pattern is
more pronounced at the higher temperature.
\par
In a typical application of the master equation, therefore, the initial processes involve
rapid equilibration between small groups of minima that are adjacent in configuration space.
This is followed by wider probability flow between larger groups of minima, and finally slow
adjustment of the population of a few minima via processes involving multiple rearrangements.

\section{Interfunnel Dynamics in LJ$_{38}$}
\label{sect:thirtyeight}

We now turn to the double-funnel energy landscape of the 38-atom Lennard-Jones
cluster, LJ$_{38}$. The potential energy is given by\cite{Jones25a}
\begin{equation}
V=\sum_{i<j}V_{ij};\hspace*{10mm}
V_{ij}=4\epsilon\left[\left(\frac{\sigma}{r_{ij}}\right)^{12}
-\left(\frac{\sigma}{r_{ij}}\right)^6\right],
\label{eq:LJ}
\end{equation}
where $\sigma$ is the pair separation at which $V_{ij}=0$, and $\epsilon$ is
the pair well depth. We will use $\sigma$ and $\epsilon$ as the units of the
quantities they measure, setting both equal to unity; the topology of the PES
is not affected by the values of these parameters.
\par
The LJ$_{38}$ cluster is too large for a complete catalogue of minima and transition states
to be obtained. However, we have previously performed a thorough characterization of the
low-energy regions of the PES using a database of 6000 minima and 8633 transition
states\cite{LJpaper,Doye99b}. The resulting disconnectivity graph\cite{LJpaper,Doye99b} clearly
showed two separated regions of configuration space, each having the character of a funnel. The
global minimum lies at the bottom of a small funnel associated with 28 minima that are
characterized by face-centred
cubic packing. The larger funnel of 446 minima leads to the lowest-energy icosahedrally
packed minimum. Since this secondary funnel accounts for a large volume of configuration space,
and because the liquid-like minima are structurally more similar to the icosahedral minima
than the fcc ones, the icosahedral funnel is expected to act as a kinetic trap for
relaxation from high-energy states to the global minimum.

\begin{figure}
\begin{center}
\epsfig{figure=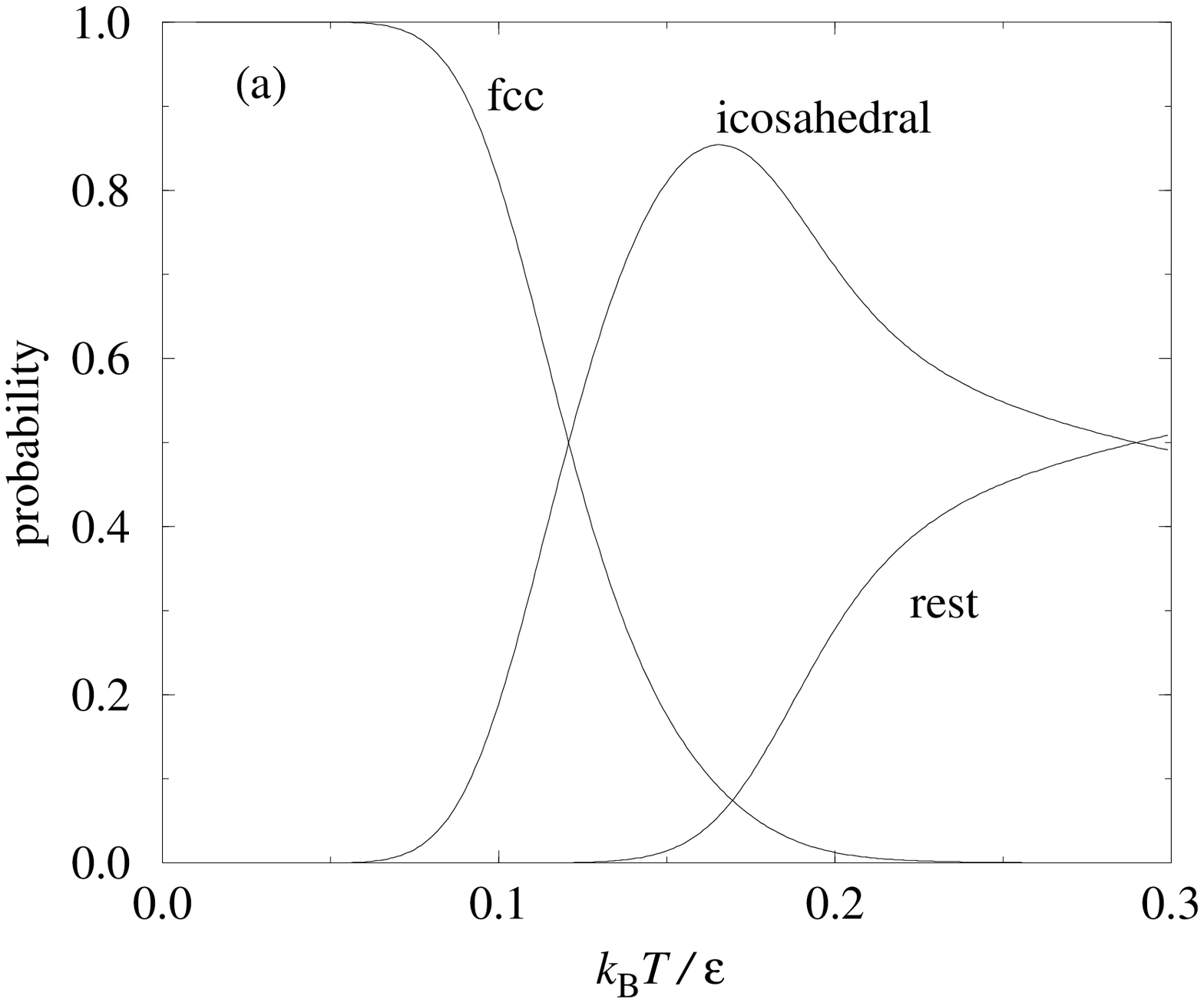,width=76mm} \\
\epsfig{figure=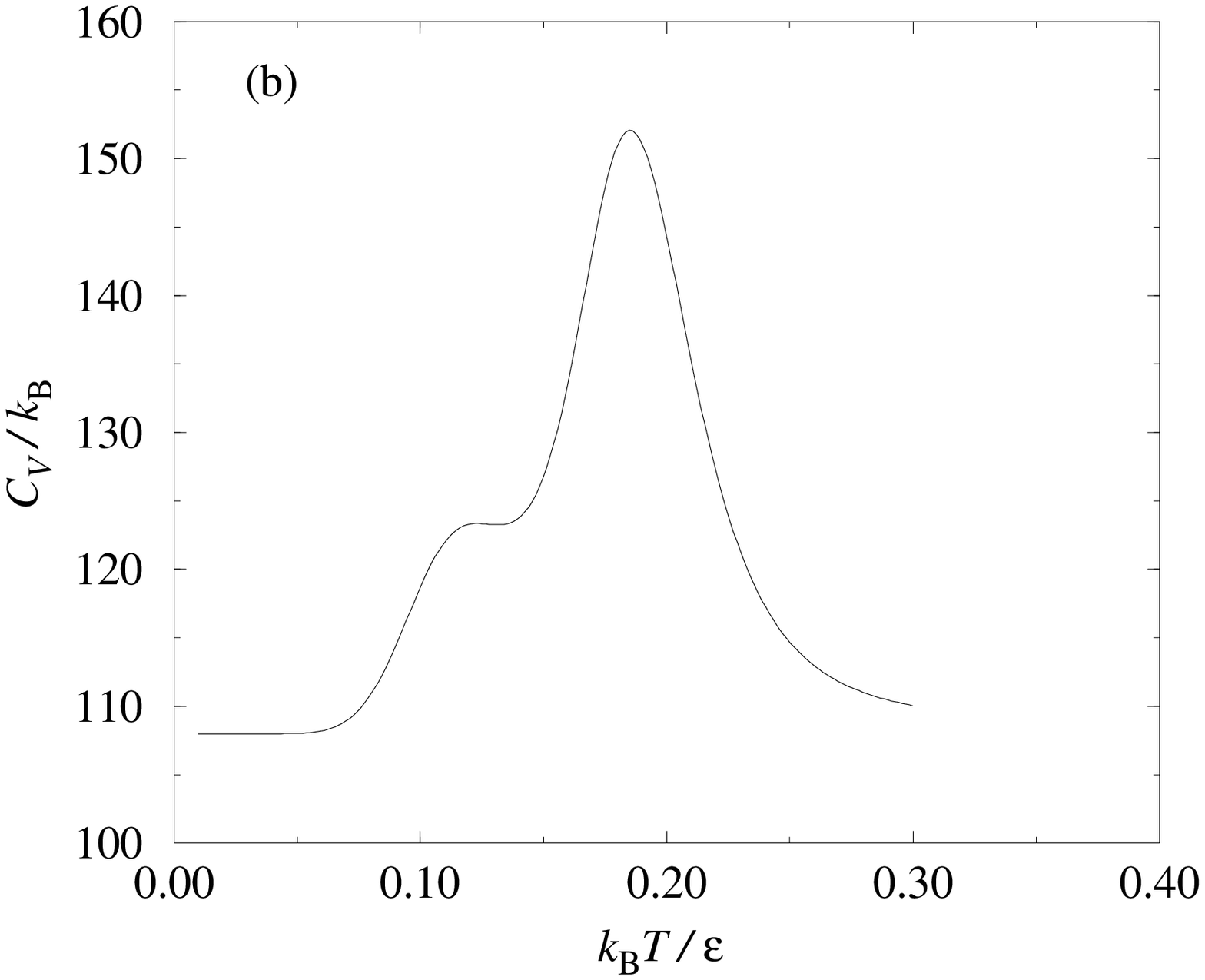,width=76mm}
\end{center}
\begin{minipage}{85mm}
\caption{
Low-temperature properties of LJ$_{38}$ calculated using the harmonic superposition
approximation. (a) Equilibrium occupation probability of the fcc and icosahedral funnels
and the rest of configuration space, and (b) the heat capacity.
}
\label{fig:lowT}
\end{minipage}
\end{figure}

By grouping together the minima in each of the two funnels, we can study not only equilibration
within the funnels as for M$_{13}$, but also the dynamics between the funnels.
Hence, we define the probabilities
\begin{equation}
P_{\rm fcc}(t)=\sum_{i\in{\rm fcc}}P_i(t) \quad\text{and}\quad
P_{\rm icos}(t)=\sum_{i\in{\rm icos}}P_i(t).
\end{equation}
The database obtained in previous work\cite{LJpaper,Doye99b} is a good representation of the
low-lying regions of the PES, but does not extend far into the liquid-like regime, and so we are
restricted to studying the dynamics at low temperatures, where the role of the liquid is
less important. This is not a serious restriction, since the time scale separation of inter-
and intrafunnel processes should be greatest at low temperature.
\par
To obtain an impression of the temperature range over which valid conclusions can be
drawn, Fig.\ \ref{fig:lowT} presents some properties calculated using the harmonic superposition
approximation and the full low-energy database of $6000$ minima and $8633$ transition
states. Fig.\ \ref{fig:lowT}a shows $P^{\rm eq}_{\rm fcc}$ and $P^{\rm eq}_{\rm icos}$ as a function
of temperature. The global minimum only dominates at very low temperatures, since the larger
number of minima and lower vibrational frequencies in the icosahedral funnel cause the
phase volume of the latter to rise rapidly. At sufficiently high
temperature the cluster should melt, and the occupation probabilities of the two funnels,
which contain predominantly solid-like structures, should approach zero. However,
the steady rise of the curve marked ``rest'' (i.e., $1-P_{\rm fcc}-P_{\rm icos}$) is interrupted
at $\kb T\approx0.2\,\epsilon$, reflecting the fact that the sample of minima does not
represent the relevant regions of the PES very well above this temperature. Figure \ref{fig:lowT}b shows
the heat capacity, derived from $C_V=(\partial U/\partial T)_V$ and Eq.\ (\ref{eq:UHO}). The small peak
at $\kb T\approx0.12\,\epsilon$ results from the transition from the fcc to icosahedral
regions of configuration space, and the main peak at $\kb T\approx0.18\,\epsilon$
signifies the melting transition. These features are largely in agreement with a more sophisticated
anharmonic superposition method, which was designed to model the thermodynamics from a representative
sample of minima\cite{Doye95a,Doye98b}. The harmonic results presented here predict a lower
peak for the melting transition, and beyond the melting temperature the heat capacity returns
to its solid-like value. This result again reveals the deficiencies of the sample of minima in the
liquid-like regime. We conclude, however, that as far as thermodynamic properties are concerned,
the present sample should be adequate for $\kb T<0.2\,\epsilon$. The corresponding limit
in the microcanonical ensemble is $E<-150\,\epsilon$, and the fcc and icosahedral funnels
have equal probability at $E=-160.5\,\epsilon$. We note that conventional
simulations would not be able to measure the quantities in Fig.\ \ref{fig:lowT} reliably at such
temperatures because the interfunnel dynamics are too slow. One of the aims of this
section is to quantify the rate of passage between the funnels.

\begin{figure}
\begin{center}
\epsfig{figure=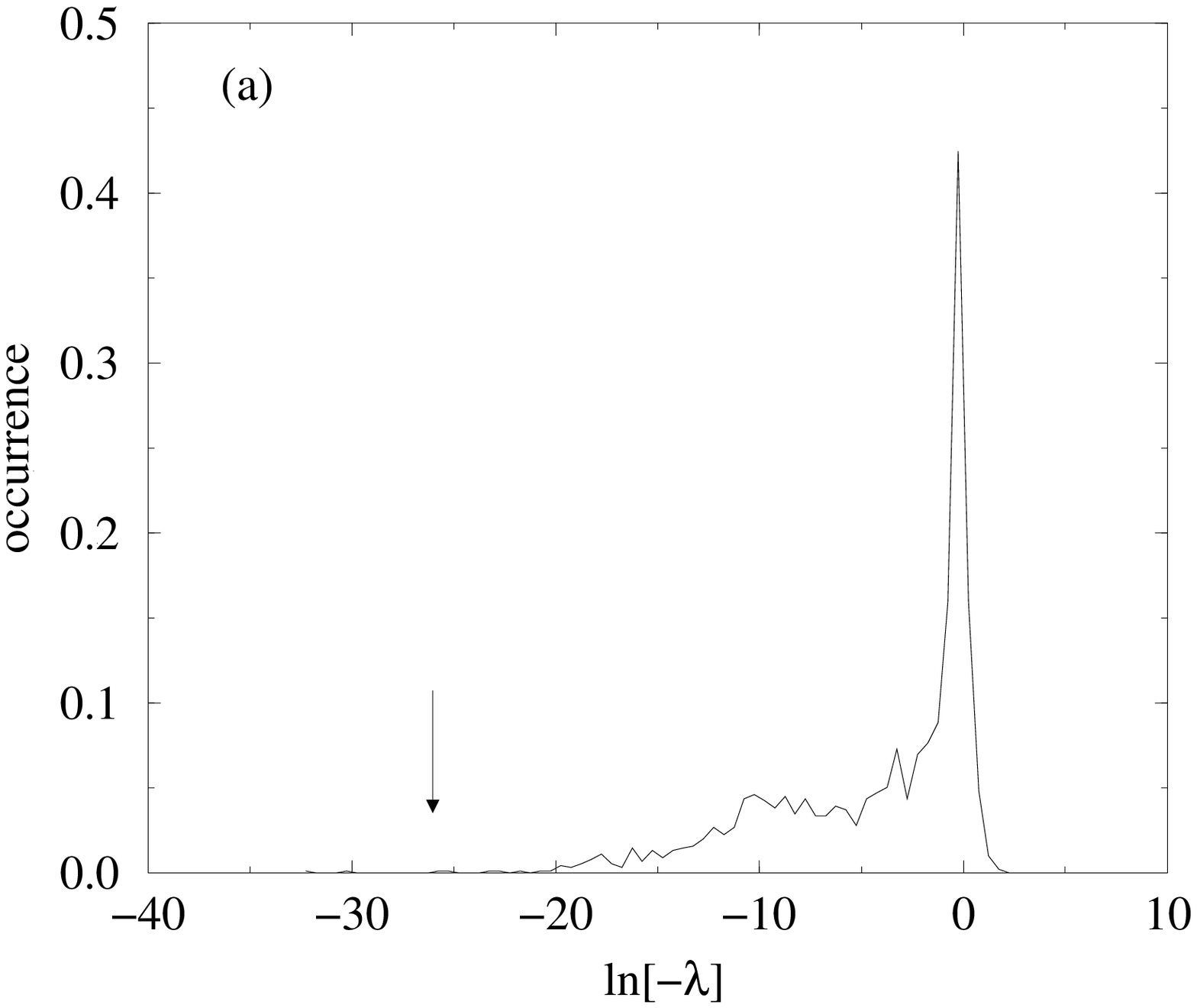,width=76mm} \\
\epsfig{figure=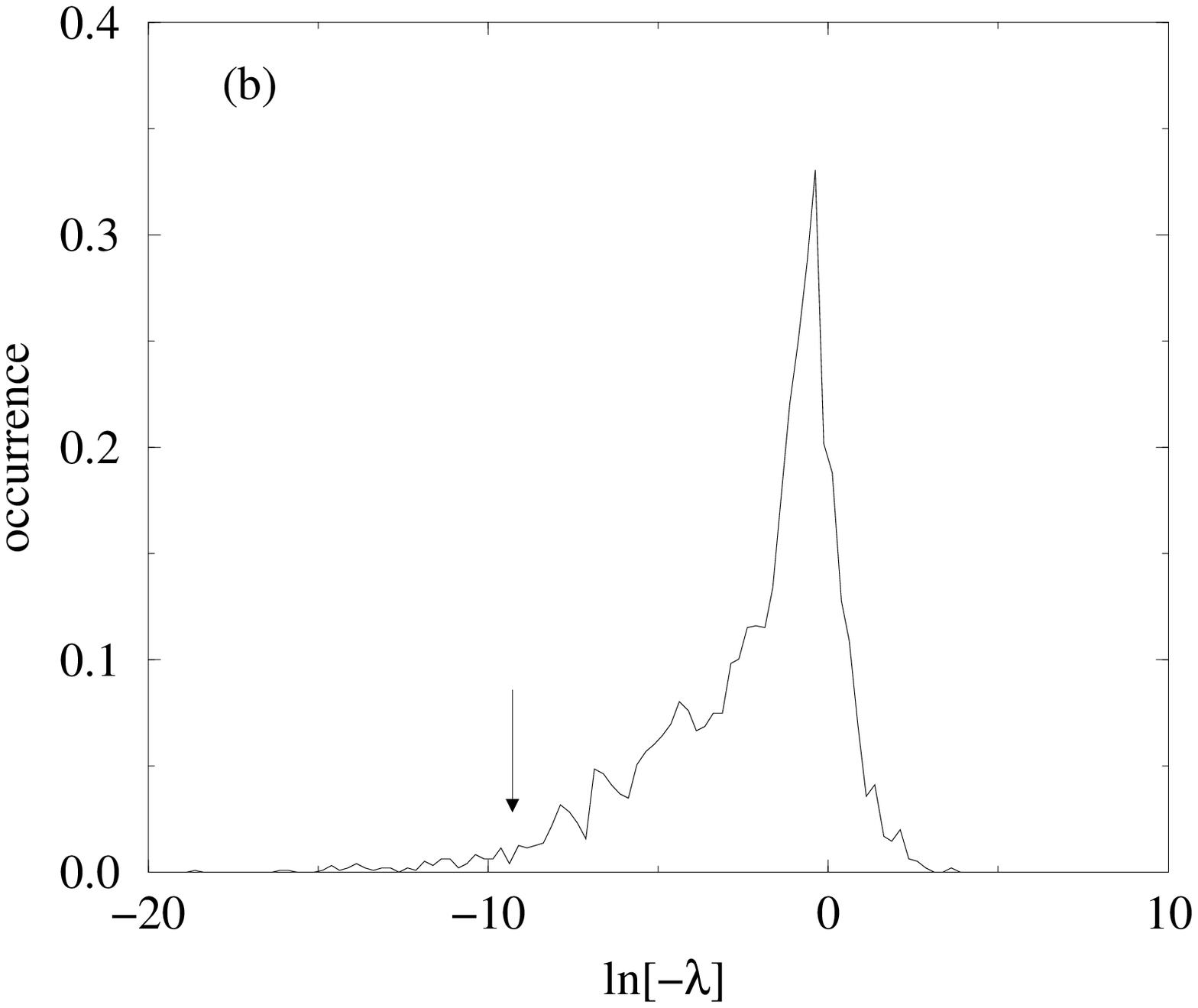,width=76mm}
\end{center}
\begin{minipage}{85mm}
\caption{
Spectra of transition matrix eigenvalues for LJ$_{38}$ in the microcanonical ensemble at
(a) $E=-160\,\epsilon$ and (b) $E=-150\,\epsilon$. In each case, the eigenvalue of
the most prominent interfunnel mode is marked by an arrow, showing the greater
separation of its time scale from the other relaxation modes at low energy.
$\lambda$ is in units of $\ljfrq$.
}
\label{fig:evspecLJ38}
\end{minipage}
\end{figure}

\subsection{Pruning the Database}
\label{sect:prune}

Because the time scale of interfunnel processes is so long at low temperatures, numerical
integration of the master equation is not feasible, so we must diagonalize the rate matrix
and use Eq.\ (\ref{eq:msolution}). However, diagonalisation routines are likely to
run into numerical problems when the matrix elements span many orders of magnitude.
Fig.\ \ref{fig:evspecLJ38} shows that lowering the temperature widens the spread of eigenvalues.
At sufficiently low temperature, a number of eigenvalues begin to
appear to be slightly positive, or the diagonalisation routines may fail altogether.
Czerminski and Elber reported similar problems\cite{Czerminski90a}, and therefore
restricted their studies to sufficiently high temperatures.
In Sec.\ \ref{sect:rangerelax} we saw that the eigenvalues at the extremes of the spectrum tend to be
associated with probability flow between a small number of minima. If these minima
are kinetically isolated, they cause the transition matrix to become nearly decomposable,
giving rise to the numerical difficulties. Since they do not participate in probability
flow we should seek to exclude them. If they are not kinetically isolated, they will
also appear in other, faster, relaxation modes.
\par
From a practical point of view, even if the transition matrix can be diagonalized without
numerical difficulty, it is worth considering whether all the minima play a significant role.
If some can be discarded without affecting the relaxation, the size of the matrix can be reduced,
saving computational effort. Because of the astronomical number of minima on the PES,
a large proportion of the minima found in a partial search are only linked to one other
minimum in the database. Within the restricted sample, these minima constitute ``dead ends'' for
probability flow. They can act as buffers, absorbing and releasing probability as it flows
towards equilibrium through the connected minimum, but they cannot act as pathways for
flow {\em between} minima or larger regions of the PES. The dead-end minima in our sample
tend to be high in energy since the search algorithm only explores connections from low-energy
minima thoroughly. When a high-energy minimum is found, the upward move is rejected and
the search continues from the original minimum, never returning to the high ones. When modelling
the probability flow from the bottom of one funnel into the other, high-energy dead-end minima are
unlikely to play an important role because their equilibrium probability is low and they do not
mediate interfunnel flow.
\par
These considerations suggest ways of pruning the database. First, dead-end minima were
identified, and were found to constitute about 70\% of the sample of $6000$ minima.
To eliminate the kinetically isolated minima at a given energy, the total outward rate
constant was calculated for each dead-end minimum, and the minimum was discarded if the
rate fell below a certain threshold. For example, at a total energy
of $E=-160\,\epsilon$, the rate constants for individual processes span the range
$10^{-41}$ to $10^0\ljfrq$. Choosing a threshold of $10^{-12}$,
which corresponds to a time scale of seconds for argon parameters, reduces the sample
of minima to $5944$. A threshold of $10^{-10}$ reduces it to $5861$. While this may be
sufficient to remove numerical difficulties in the diagonalisation procedure, the
matrices are still rather large.
\par
Trimming the sample on the basis of equilibrium probabilities reduced the
number of minima more dramatically. At $E=-160\,\epsilon$, discarding dead-end minima whose
equilibrium probability was less than $10^{-10}$ left just $1825$ minima, and a threshold
of $10^{-8}$ left $1782$. Clearly, the number of minima removed by such a method
decreases with increasing energy, since higher-energy states then become more populated.
We will gauge the effect of pruning the database by examining the sensitivity of the
results to the choice of the threshold.

\subsection{Interfunnel Rate Constants}

What is the rate of crossing between the fcc and icosahedral funnels? 
We have previously shown that
interconversion of the fcc and icosahedral minima is a multiple-step
process\cite{LJpaper}---the lowest-energy path in our sample involves
13 successive rearrangements---but let us consider the overall scheme
\begin{equation}
\text{fcc}\rightleftharpoons\text{icos},
\label{eq:scheme}
\end{equation}
with ``forward'' and ``reverse'' rate constants $k_+$ and $k_-$.
The rate of change of the occupation probability of the fcc funnel is accordingly
\begin{equation}
\frac{dP_{\rm fcc}(t)}{dt}=-k_+P_{\rm fcc}(t)+k_-P_{\rm icos}(t).
\label{eq:fccprob}
\end{equation}
From Fig.\ \ref{fig:lowT}, we see that at sufficiently low temperature only the
two funnels are significantly occupied at equilibrium, as opposed to minima
associated with the liquid-like state. Assuming that
this is also the case away from equilibrium, provided that the initial
probability is itself confined to the funnels, we can write
\begin{equation}
P_{\rm fcc}(t)+P_{\rm icos}(t)=1.
\label{eq:sumprob}
\end{equation}
Using Eq.\ (\ref{eq:sumprob}) and the equilibrium relationship
\begin{equation}
k_+P^{\rm eq}_{\rm fcc}=k_-P^{\rm eq}_{\rm icos},
\label{eq:eqm}
\end{equation}
integration of Eq.\ (\ref{eq:fccprob}) gives the basic result of first order kinetics for
a two-state model:
\begin{equation}
\ln\left[\frac{P_{\rm fcc}(t)-P^{\rm eq}_{\rm fcc}}
{P_{\rm fcc}(0)-P^{\rm eq}_{\rm fcc}}\right]
=-(k_+ + k_-)t.
\label{eq:twostate}
\end{equation}
\par
\begin{figure}
\begin{center}
\epsfig{figure=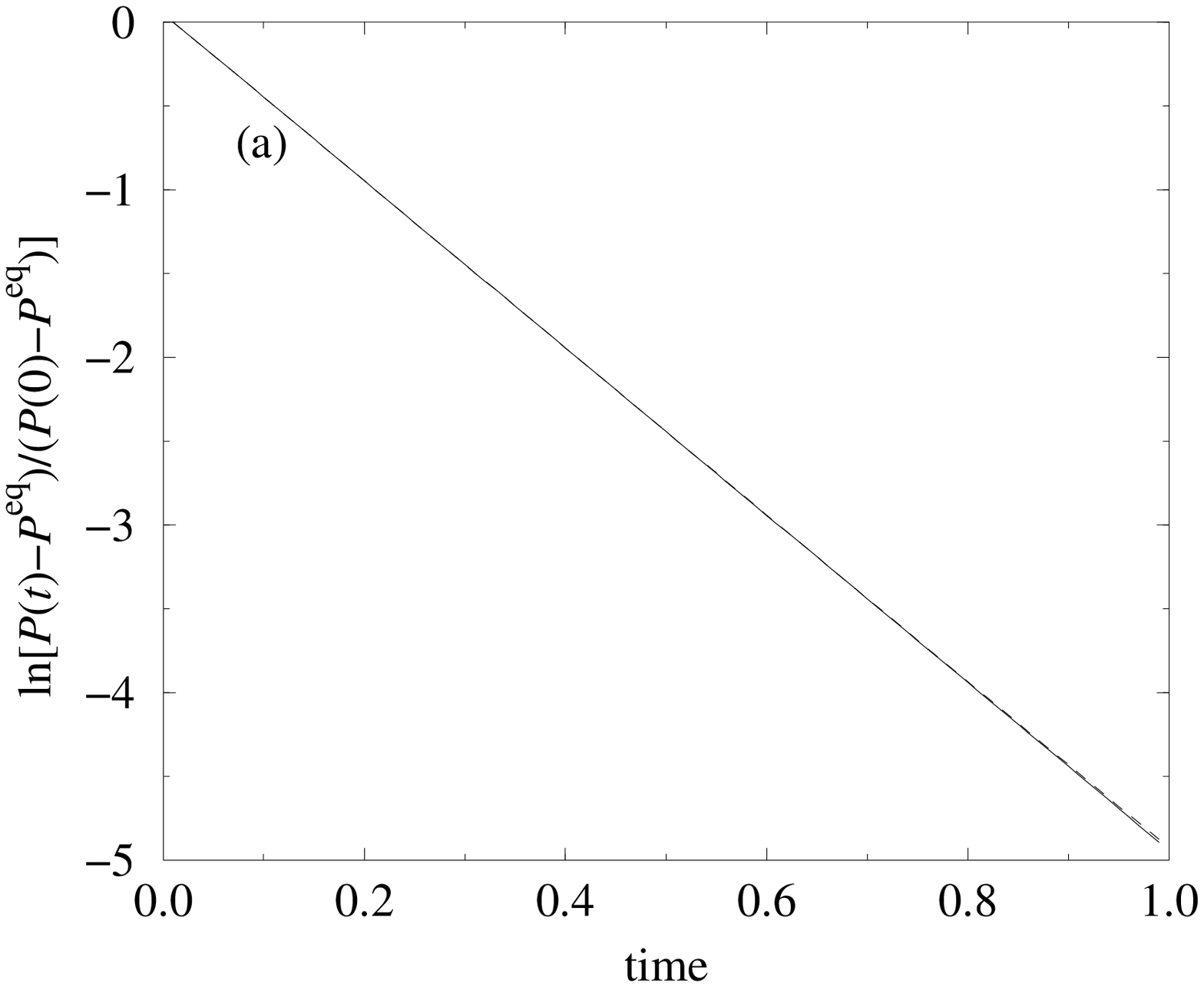,width=76mm} \\
\epsfig{figure=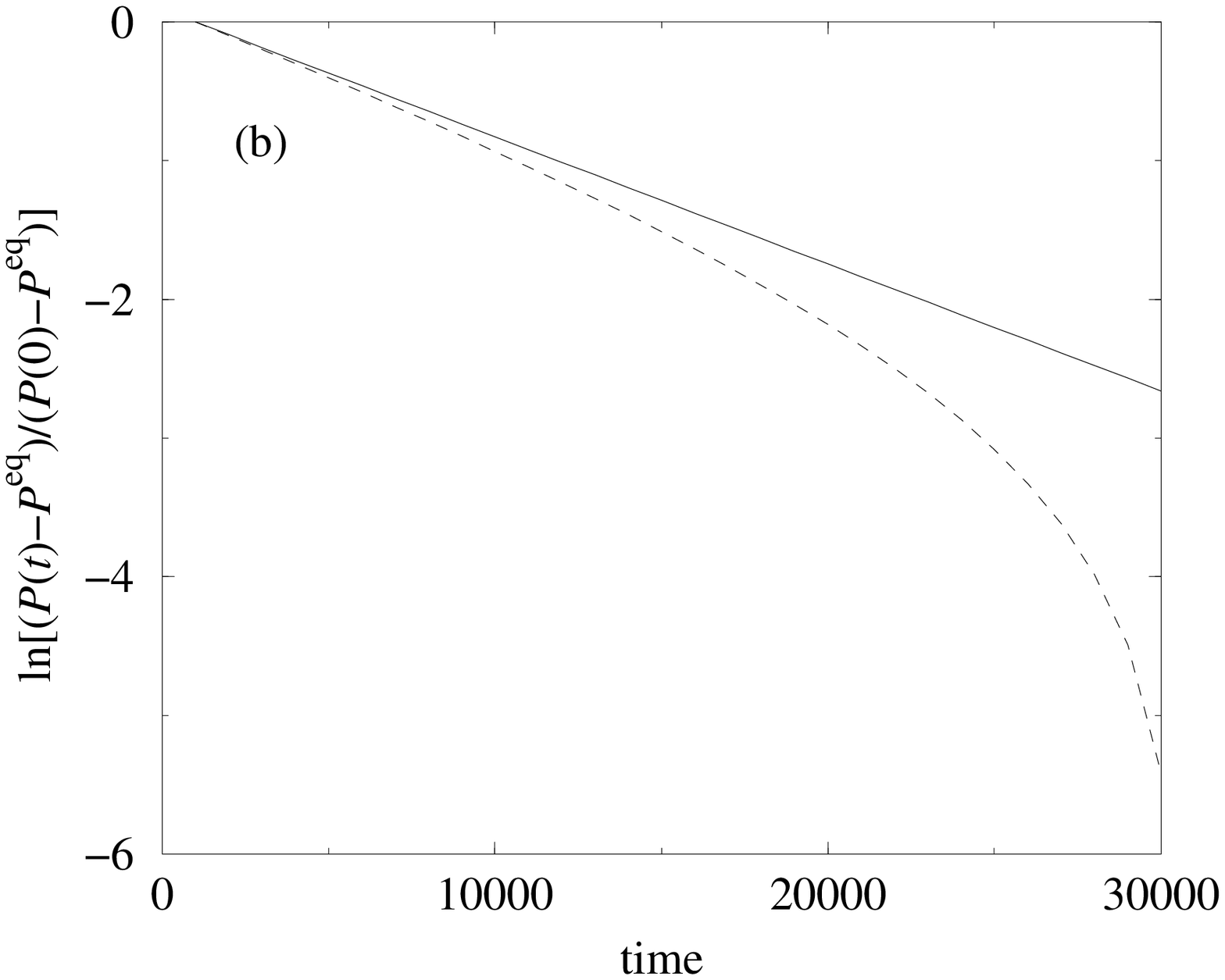,width=76mm}
\end{center}
\begin{minipage}{85mm}
\caption{
Plots of Eq.\ (\ref{eq:twostate}) for LJ$_{38}$ in the microcanonical ensemble with the initial
probability of the global minimum set to unity. Solid lines are Eq.\ (\ref{eq:twostate}) for
the fcc funnel, and dashed lines are the equivalent for the icosahedral funnel.
(a) $E=-160\,\epsilon$, (b) $E=-150\,\epsilon$. In (a) the lines coincide.
The units of time are (a) $10^{12}\ljtime$, and (b) $\ljtime$.
}
\label{fig:frates}
\end{minipage}
\end{figure}

Figure \ref{fig:frates}a shows plots of Eq.\ (\ref{eq:twostate}) and the analogous expression for
$P_{\rm icos}(t)$ in the microcanonical ensemble at $E=-160\,\epsilon$, starting
from the global minimum. The plots were obtained from the analytic solution of
the master equation after removing all dead-end minima with an equilibrium probability
of less than $10^{-8}$, as described in Sec.\ \ref{sect:prune}. The two lines are straight and
coincide, and the slope yields $k_++k_-=4.99\times10^{-12}\ljfrq$.
This value closely matches one of the eigenvalues of the transition matrix,
$|\lambda_4|=4.98\times10^{-12}\ljfrq$,
suggesting that the corresponding eigenvector describes flow between the two funnels.
The ``net flow index'' into or out of a funnel F produced by relaxation mode $i$ of the 
master equation can be obtained by summing the components of eigenvector $i$ that correspond
to the minima belonging to F\cite{Kunz95a}:
\begin{equation}
f^{\rm F}_i=\sum_{j\in{\rm F}}\tilde u^{(i)}_j \sqrt{P^{\rm eq}_j}.
\label{eq:flow}
\end{equation}
If mode $i$ represents probability flow between the funnels, then $f^{\rm fcc}_i$ and
$f^{\rm icos}_i$ will be larger in magnitude than for other eigenvectors, and will
have opposite signs, so that an increasing contribution is made to one funnel and
a decreasing contribution to the other, depending on the initial probability vector
[see Eq.\ (\ref{eq:msolution})]. At $E=-160\,\epsilon$, for the mode whose eigenvalue matches
$-(k_++k_-)$, we find $f^{\rm fcc}_4=0.495$ and $f^{\rm icos}_4=-0.495$. The next
largest net flow index for the fcc funnel was $f^{\rm fcc}_{93}=3.97\times10^{-4}$
(with a corresponding index for the icosahedral funnel of
$f^{\rm icos}_{93}=9.94\times10^{-7}$) and that for the icosahedral funnel was
$f^{\rm icos}_9=-0.143$ (with a corresponding index for the fcc funnel of
$f^{\rm fcc}_9=4.38\times10^{-5}$). This result unambiguously identifies the fourth
mode with interfunnel relaxation at this energy.

To test the effect of having pruned the database, we lowered the threshold for
removal of minima from an equilibrium probability of $10^{-8}$ to $10^{-12}$,
resulting in a larger remaining sample of $2289$ minima. The net flow index picked
out relaxation mode 17, whose eigenvalue was $\lambda_{17}=-4.98\times10^{-12}\ljfrq$,
i.e., the same as was obtained with the higher threshold. Replotting Fig.\ \ref{fig:frates}a
also yielded the same result as before. We note that, even though this new
sample is less than half the size of the full database, the lowest eigenvalues
already clash with the precision of the ``zero'' eigenvalue. Hence, some pruning is
essential for numerical tractability and desirable for computational speed, and it
does not affect the result in this case.
\par
At low energy, the interfunnel mode is easily identified. As the energy is
raised and more processes become ``unfrozen'', the distinction is somewhat less
clear. Figure \ref{fig:frates}b shows Eq.\ (\ref{eq:twostate}) starting from the global minimum at
$E=-150\,\epsilon$. The database was pruned using an equilibrium probability
threshold of $10^{-8}$ and contained $3789$ minima.
Whilst the decay of $P_{\rm fcc}(t)$ obeys the linearized
relationship, the evolution of $P_{\rm icos}(t)$ deviates from it increasingly as
time progresses. This deviation is partly because minima outside the two funnels have
non-negligible populations, so Eq.\ (\ref{eq:sumprob}) does not hold, and also because
$P_{\rm icos}(t)$ does not rise monotonically to its equilibrium value, but
overshoots slightly and then decays. The net flow index picks out
$\lambda_{95}=-9.19\times10^{-5}\ljfrq$ with $f^{\rm fcc}_{95}=0.0859$ and
$f^{\rm icos}_{95}=-0.0686$. These values are considerably smaller than those obtained
at $E=-150\,\epsilon$. Although other modes may have a higher flow index for
one funnel, the value for the other funnel is then either much smaller (indicating
that the mode describes flow between one funnel and the non-funnel states), or
of the same sign (indicating that flow is not {\em between} the funnels).
\par
The slope of the solid line in Fig.\ \ref{fig:frates}b is $-9.35\times10^{-5}\ljfrq$,
which is not far from $\lambda_{95}$, but is actually closer to
$\lambda_{96}=-9.39\times10^{-5}\ljfrq$. However, mode 96 is only weakly
interfunnel: $f^{\rm fcc}_{96}=1.91\times10^{-4}$ and
$f^{\rm icos}_{96}=-5.36\times10^{-4}$. At relatively high energies,
where minima outside the funnels come into play and the simplified scheme
of Eq.\ (\ref{eq:scheme}) breaks down, the net flow index therefore still provides a convenient
way of identifying the most important interfunnel relaxation mode and
extracting the quantity $(k_++k_-)$. This method also has the advantage
that it does not require evaluation of the master equation solution itself.
\par
The equilibrium relationship Eq.\ (\ref{eq:eqm}) allows the separate rate constants
$k_+$ and $k_-$ for interfunnel flow to be obtained from the eigenvalue.
Table \ref{tab:flow} shows $k_+$ and $k_-$ as a function of temperature in the canonical
ensemble. At each temperature, the database was pruned using an equilibrium probability
threshold of $10^{-8}$, and the table shows how many of the full sample of $6000$
minima and $8633$ transition states remained. At the lowest two temperatures, the
pruning procedure removed all dead-end minima. The table also shows that the
net flow indices for the interfunnel mode decrease in magnitude at the higher temperatures.
This reduction is a result of the increasing involvement of higher-energy and non-funnel
minima. Over the course of doubling the temperature, $k_+$ changes by ten orders
of magnitude. For argon parameters, the span of time scales is hundreds of nanoseconds
to an hour.

\end{multicols}
\begin{center}
\begin{table}
\begin{minipage}{175mm}
\caption{
Interfunnel rate constants for LJ$_{38}$ in the canonical ensemble. $n_{\rm min}'$
and $n_{\rm ts}'$ are the numbers of minima and transition states remaining in the
database after discarding minima with an equilibrium probability of less than
$10^{-8}$. $f^{\rm fcc}$, $f^{\rm icos}$ and $\lambda$ are the net flow indices
and eigenvalue of the interfunnel mode. $\lambda$, $k_+$, and $k_-$
are tabulated in units of $\ljfrq$.
}
\label{tab:flow}
\end{minipage}
\begin{minipage}{140mm}
\begin{tabular}{ccclccll}
$\kb T/\epsilon$ & $n_{\rm min}'$ & $n_{\rm ts}'$ & \multicolumn{1}{c}{$\lambda$} &
$f^{\rm fcc}$ & $f^{\rm icos}$ & \multicolumn{1}{c}{$k_+$} & \multicolumn{1}{c}{$k_-$} \\
\hline
0.09 & 1770 & 4371 & $-8.50\times10^{-15}$ & 0.282 & -0.281 & $7.36\times10^{-16}$ & $7.77\times10^{-15}$ \\
0.10 & 1770 & 4371 & $-3.96\times10^{-13}$ & 0.399 & -0.392 & $7.48\times10^{-14}$ & $3.22\times10^{-13}$ \\
0.11 & 1783 & 4384 & $-9.74\times10^{-12}$ & 0.471 & -0.470 & $3.22\times10^{-12}$ & $6.52\times10^{-12}$ \\
0.12 & 1809 & 4410 & $-1.52\times10^{-10}$ & 0.500 & -0.499 & $7.39\times10^{-11}$ & $7.80\times10^{-11}$ \\
0.13 & 1851 & 4453 & $-1.66\times10^{-9}$  & 0.483 & -0.482 & $1.06\times10^{-9}$  & $5.97\times10^{-10}$ \\
0.14 & 1978 & 4583 & $-1.35\times10^{-8}$  & 0.438 & -0.435 & $1.00\times10^{-8}$  & $3.53\times10^{-9}$  \\
0.15 & 2264 & 4872 & $-8.64\times10^{-8}$  & 0.381 & -0.374 & $7.09\times10^{-8}$  & $1.54\times10^{-8}$  \\
0.16 & 2620 & 5232 & $-4.46\times10^{-7}$  & 0.320 & -0.306 & $3.92\times10^{-7}$  & $5.36\times10^{-8}$  \\
0.17 & 2985 & 5599 & $-1.92\times10^{-6}$  & 0.259 & -0.229 & $1.77\times10^{-6}$  & $1.52\times10^{-7}$  \\
0.18 & 3363 & 5981 & $-7.13\times10^{-6}$  & 0.202 & -0.196 & $6.77\times10^{-6}$  & $3.57\times10^{-7}$  \\
\end{tabular}
\end{minipage}
\end{table}
\end{center}
\begin{multicols}{2}

\begin{figure}
\begin{center}
\epsfig{figure=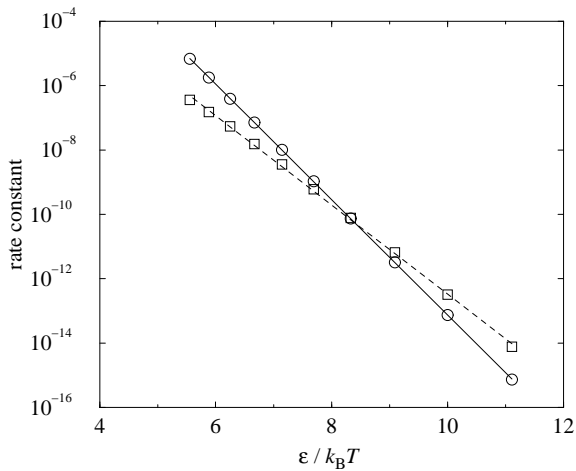,width=76mm}
\end{center}
\begin{minipage}{85mm}
\caption{
Arrhenius plot of $k_+$ (circles) and $k_-$ (squares) for the interfunnel dynamics
of LJ$_{38}$. The lines are fits to the form $k=A\exp(-E_{\rm a}/\kb T)$.
The units of the rate constant are $\ljfrq$.
}
\label{fig:Arrhenius}
\end{minipage}
\end{figure}

Figure \ref{fig:Arrhenius} shows that, over the temperature range in Table \ref{tab:flow}, $k_+$ and $k_-$
obey an Arrhenius temperature dependence law. Only very slight curvature is visible in
the $k_-$ results. Fitting to the form $k=A\exp(-E_{\rm a}/\kb T)$ gives the pre-exponential
factors and activation energies for the forward and reverse processes:
\begin{eqnarray*}
k_+:\ \text{fcc} \rightarrow \text{icos} & \qquad A=11.1\ljfrq & \qquad E_{\rm a}=4.12\,\epsilon \\
k_-:\ \text{icos} \rightarrow \text{fcc} & \qquad A=3.18\ljfrq & \qquad E_{\rm a}=3.19\,\epsilon
\end{eqnarray*}

Interestingly, the effective activation energy for $\text{fcc}\to\text{icos}$ is close to
the overall barrier on the lowest-energy path between the funnels, starting from the
global minimum\cite{LJpaper}, which is $4.22\,\epsilon$. This result suggests that the pathways
passing over the highest-energy transition state on the lowest-energy pathway determine the
interfunnel dynamics. $E_{\rm a}$ for $\text{icos}\to\text{fcc}$, however, is significantly
lower than the overall potential barrier for the reverse process, which is $3.54\,\epsilon$,
starting from the lowest-energy icosahedral minimum. The
discrepancy can be attributed to the fact that more than one minimum in the icosahedral funnel
is substantially occupied. Hence, the effective barrier from the icosahedral funnel should not
be measured from the lowest-energy minimum in the icosahedral funnel, but with respect to
a weighted average of the minimum energies, $\sum_{i\in{\rm icos}}P_i V_i$. The slight
curvature in the Arrhenius plot for $k_-$ is a result of the temperature dependence of this
average.
\par
An alternative method for computing the rates for interfunnel conversion would be to calculate
first the free energy barrier between the two funnels, and then the transmission coefficient
for passage over this barrier\cite{Chandler}. The free energy barriers have been calculated
for LJ$_{38}$ and were found to decrease as the temperature is increased towards that required
for melting. Therefore, if this method is also to show an Arrhenius behaviour, the temperature
dependence of the transmission coefficient must compensate that of the free energy barriers.
\par
We have seen that the interfunnel rates drop dramatically as the temperature is lowered.
At the same time, the increasing net flow indices show that the corresponding relaxation
modes become more distinct from other processes. These features concur with
Stillinger's interpretation of $\alpha$ processes in fragile liquids\cite{Stillinger95a}.
In this picture, $\beta$ processes are faster and more localized in configuration space,
whereas the $\alpha$ processes, which become relatively slow at low temperatures,
are thought to occur between ``craters'' (a similar concept to funnels)
on the energy landscape. These differences
give rise to a bifurcation of time scales, which is visible for LJ$_{38}$ in the eigenvalue
spectra of Fig.\ \ref{fig:evspecLJ38}.
However, $\alpha$ processes have been observed to have a non-Arrhenius temperature
dependence, in contrast with the results of Fig.\ \ref{fig:Arrhenius}. The Arrhenius behaviour
may be a genuine feature of the dynamics of the LJ$_{38}$ cluster, but could also be a
limitation of the present approach. In particular, our incomplete database contains only a
limited number of pathways between the two funnels, and competition between such
paths would give rise to non-Arrhenius behaviour.
We can study only a limited temperature range with our database, and the modelling
of individual processes might not be sufficiently sophisticated. We note that Angelani
et al.~also observed unexpected Arrhenius behaviour in their master equation study of
a fragile glass former\cite{Angelani98a,Angelani99a}.
 
\subsection{Equilibration}

The progress of the probability vector towards equilibrium can be visualized using
equilibration graphs\cite{Doye96b,Sibani93a,Schon96a,Schon96b}. Such a graph
has a time axis, on which lines denote a group of states in
local equilibrium with each other. Nodes join lines at the time that the
corresponding groups first come into equilibrium, until there is just one group and
the whole system has equilibrated. We define the
time that minima $i$ and $j$ come into equilibrium as the smallest value of $t$ after which
\begin{equation}
\frac{\left|P_i(t)P^{\rm eq}_j-P_j(t)P^{\rm eq}_i\right|}
{\sqrt{P_i(t)P_j(t)P^{\rm eq}_iP^{\rm eq}_j}}
\leq\xi
\label{eq:eqtime}
\end{equation}
is always satisfied, and in the present work we set $\xi=0.01$.
\par
\begin{figure}
\begin{center}
\epsfig{figure=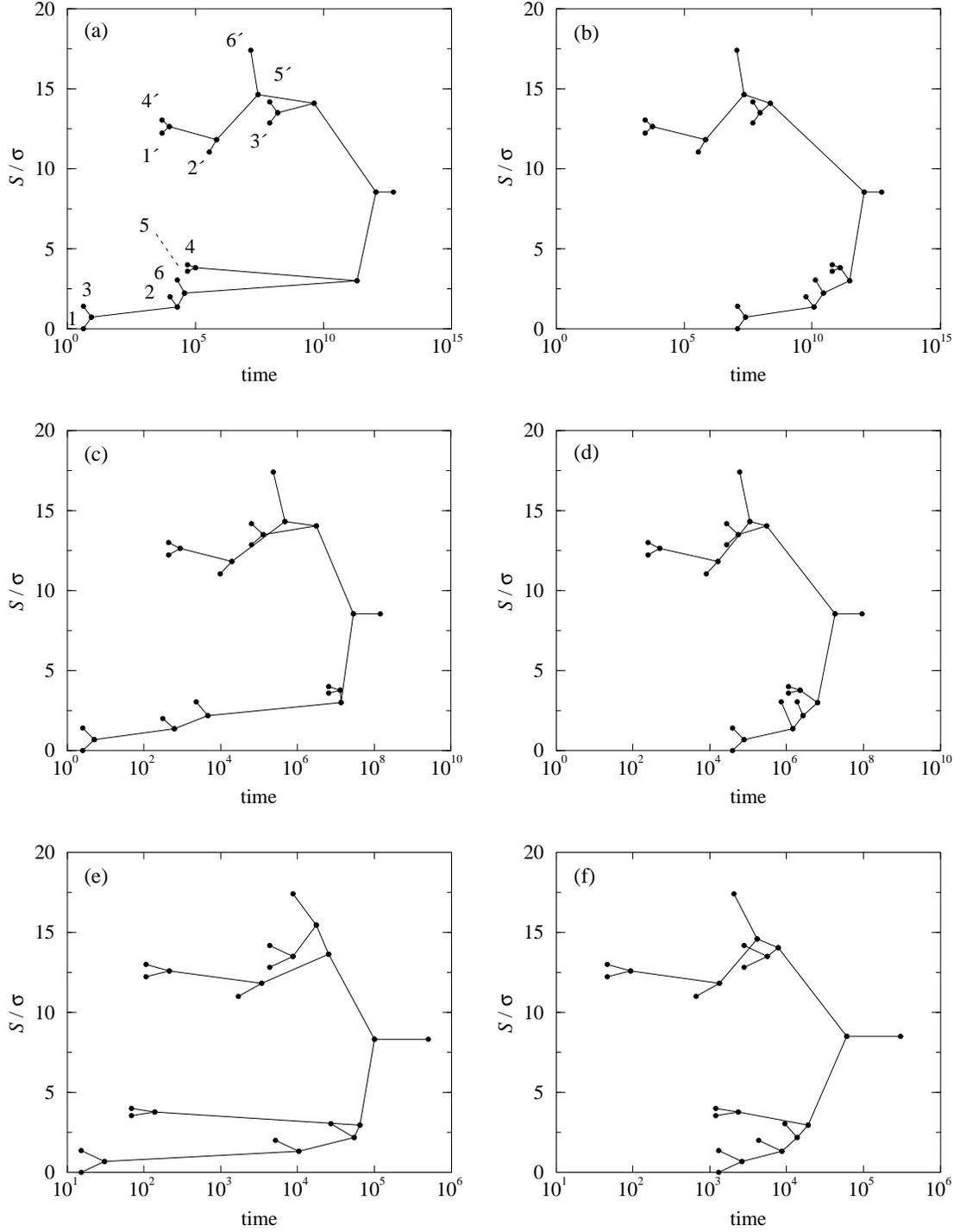,width=150mm}
\begin{minipage}{150mm}
\caption{
Equilibration graphs for LJ$_{38}$ in the microcanonical ensemble at three energies:
(a) and (b) $-160\,\epsilon$, (c) and (d) $-155\,\epsilon$, (e) and (f) $-150\,\epsilon$.
In each row, the left-hand graph is for the global minimum (labelled 1) having an initial
probability of 1, and the right-hand graph is for the lowest-energy icosahedral minimum
(labelled $3'$) having an initial probability of 1.
Lines representing individual minima commence at a vertical position corresponding
to the shortest integrated path length, $S^{\rm gm}$, to the global minimum, and an
arbitrary horizontal position. Nodes join lines when the corresponding states first
come into equilibrium. In (a), unprimed numbers indicate minima in the fcc funnel,
and primed numbers indicate minima in the icosahedral funnel.
The time is in units of $\ljtime$.
}
\label{fig:eqtrees}
\end{minipage}
\end{center}
\end{figure}

Figure \ref{fig:eqtrees} shows equilibration graphs for the six minima in each of the two funnels
of LJ$_{38}$ that have the greatest equilibrium probability at $E=-150\,\epsilon$. They
are numbered in order of increasing probability within each funnel. The initial vertical
position of each minimum is taken as the integrated
path length, $S^{\rm gm}$, along the shortest path to the global minimum, making the
two groups clearly distinguishable on this axis. Three microcanonical energies are
plotted, spanning the range of applicability of our database. The sample was pruned
using an equilibrium threshold of $10^{-8}$ at each energy. The evolution of two
initial probability vectors was considered: in the left-hand graphs, the initial probability
of the global minimum is unity, and in the right-hand graphs the probability commences
in the lowest-energy icosahedral minimum.
\par
In each of the six graphs, the minima within a funnel come into equilibrium with each
other before the separate funnels do so. This result explicitly demonstrates the
longer time scale of the interfunnel dynamics. The order of equilibration within
each funnel is the same at all three energies studied, irrespective of the funnel in
which the probability is initiated. There is a small exception in graph (e), where minimum
$6'$ in the icosahedral funnel equilibrates with $3'$ and $5'$ before it joins $1'$, $2'$
and $4'$. Interestingly, the lowest-energy icosahedral minimum, $3'$, is one of the last
minima to reach equilibrium within the icosahedral funnel, presumably because of the
high barriers surrounding it\cite{LJpaper}.
\par
The equilibration of the fcc funnel is much more sensitive to the initial probability
than that of the icosahedral funnel, in spite of the fact that the rate constants,
$k_+$ and $k_-$, are roughly equal at $E=-160\,\epsilon$. This difference in behaviour
of the two funnels arises from the fact that the absolute probabilities of the minima
in the fcc funnel, other than the global minimum itself, are several orders of magnitude
smaller than those of the minima in the icosahedral funnel. Hence, small changes in
probability due to transient flows easily disturb the equilibrium between minima in
the fcc funnel, since distance from equilibrium is measured relative to the final
probability by the left-hand side of inequality (\ref{eq:eqtime}). When the cluster
starts in the global minimum, the probability of the fcc minima decreases monotonically,
and the minima within the funnel can equilibrate with each other rapidly. When the probability
is initialized in the lowest-energy icosahedral minimum, however, the influx of probability to the
fcc funnel must pass through the minima within the funnel on its way to the global minimum,
and the transients must settle down almost completely before equilibrium is
permanently established. This effect is reflected by the shift of the equilibration
nodes of the fcc funnel to later times as one goes from the left-hand to right-hand
side equilibration graphs in Fig.\ \ref{fig:eqtrees}.
\par
\begin{figure}
\epsfig{figure=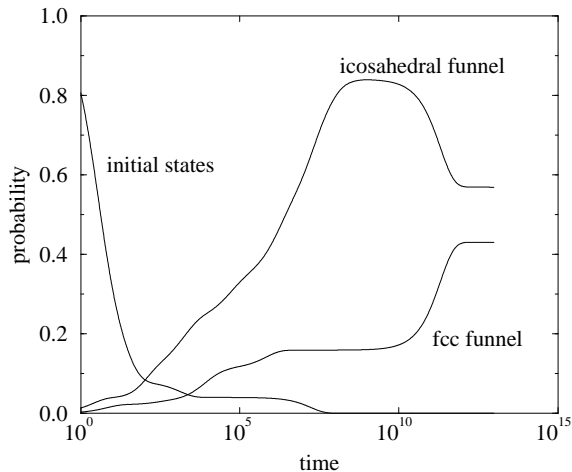,width=76mm}
\begin{minipage}{85mm}
\caption{
Relaxation of LJ$_{38}$ from high-energy minima at a total energy of $-160\,\epsilon$,
showing the fast and slow contributions to the final probability of the fcc
funnel. The time is in units of $\ljtime$.
}
\label{fig:fastslow}
\end{minipage}
\end{figure}

At sufficiently low energy, the global potential energy minimum must be the most populated
state at equilibrium. However, we have seen that there is a kinetic bottleneck to entering
its funnel. Hence, if the cluster is prepared in a liquid-like state, it is most likely to
collapse into the icosahedral funnel, which is larger and structurally more similar to the
liquid, even though it is not the equilibrium state. Over a sufficiently long time, the
cluster must then convert to the global minimum. Our master equation model shows this behaviour
clearly. Starting from a uniform distribution amongst the 25 highest-energy minima in the
sample, $P_{\rm fcc}(t)$ and $P_{\rm icos}(t)$ were monitored as the system evolved towards
equilibrium at the low energy of $E=-160\,\epsilon$. The results are shown in Fig.\ \ref{fig:fastslow}.
\par
The initial states decay rapidly into other non-funnel states, and the two funnels experience
a slow increase in population. Although some probability enters the fcc funnel, it reaches a plateau
while the population of the icosahedral funnel continues to grow. This growth reaches a maximum
before eventually decaying towards its equilibrium value, as the probability trickles into the global
minimum. The icosahedral funnel acts as a kinetic trap, and only releases the cluster into
the global minimum on a long time scale. We note that direct simulation of the trapping effect
by standard MD would therefore be highly problematic.

\end{multicols}
\begin{multicols}{2}

Although Fig.\ \ref{fig:fastslow} unambiguously demonstrates the separate fast and slow contributions to
the relaxation, the precise partitioning between the two funnels at the plateau stage can depend
on the initial probability distribution. The distribution chosen here is rather artificial because
our sample of minima does not extend into the liquid-like range. If we could release the system
from a high-temperature liquid-like state, the fraction of probability flowing into the icosahedral
funnel would probably be even larger because of the greater structural similarity of the liquid-like
structures with the icosahedral rather than fcc minima. Although the fraction of fcc minima is already
small in our database, it would be much smaller in a more comprehensive sample.
\par
Two-stage dynamics have been observed experimentally
in the folding of hen egg lysozyme, in which two routes to the native state have been postulated:
one fast and direct, the other passing via partially folded conformations which act as kinetic
traps and reorganize to the native state only slowly\cite{Kiefhaber95a}. Conversely, the
protein plasminogen activator inhibitor 1 rapidly folds to the active state, but converts to
an inactive form on a much slower time scale\cite{Baker94a}, implying that the active state
is not the global free energy minimum but is only metastable, like the icosahedral minima in
LJ$_{38}$ at low energy. Similarly, human prion protein has two long-lived forms, whose
relative stability can be adjusted by varying the pH. The time scale for conversion to the
more stable form is of the order of days\cite{Jackson99a}.

\section{Summary}
\label{sect:summary}

We have applied the master equation to the structural databases
for M$_{13}$ and LJ$_{38}$ derived in previous work\cite{Miller99a,Doye99b} to model
relaxation processes in these atomic clusters. This approach can be applied to time
scales far longer than those accessible by direct simulation, and describes the behaviour
of an equilibrating ensemble without the need to average over separate trajectories.
\par
The harmonic approximation for the density of states of individual minima and
transition states provides a simple but physically clear basis for calculating
equilibrium properties and rate constants. Provided it is not applied at excessively
high temperatures, it gives a qualitatively useful description of the thermodynamic
and dynamic properties of the energy landscape.
\par
As we predicted\cite{Miller99a},
relaxation to the global minimum is slower when the range of the potential is
shorter. An optimal temperature for this relaxation is obtained by a compromise between
the decreasing rates at low temperatures and the decreasing thermodynamic driving
force at high temperatures. When the range of the potential is long, the cluster
exhibits a wide temperature window over which relaxation is quite efficient. In
contrast, when the range is short, small deviations from the optimal temperature
hinder the rate appreciably.
\par
Although the relaxation profiles of the total energy at fixed temperature do not
appear to be well described by any of the commonly used empirical forms, the
temperature dependence of the mean relaxation time followed an Arrhenius law
for $\rho=14$ and a Vogel--Tammann--Fulcher law at $\rho=4$. These results again
reflect the greater uniformity of the short-range PES that was deduced in the
landscape analysis.
\par
Application of the analytic solution of the master equation to the low-energy
database of LJ$_{38}$ required the removal of unimportant minima and transition
states from the sample. ``Dead-end'' minima were removed if their equilibrium
probability fell below a low threshold at the temperature of interest. The
relaxation modes of the resulting database were analysed using a flow index
to extract the rate of passage between the two funnels on the energy landscape
at low temperatures. The equilibration patterns within and between the funnels
clearly revealed the double-funnel structure. High-energy distributions relaxed
preferentially into the secondary funnel of icosahedral minima rather than
the close-packed funnel surrounding the global minimum. This behaviour stems
from the greater structural similarity of the liquid to the icosahedral
minima, which is reflected in the patterns of connectivity on the PES. Eventually,
the cluster escaped from this kinetic trap into the global minimum, which
is thermodynamically favoured at sufficiently low temperature.

\section*{Acknowledgements}

D.J.W.\ is grateful to the Royal Society
and M.A.M.\ to the Engineering and Physical Sciences Research Council
for financial support.
J.P.K.D. is the Sir Alan Wilson Research Fellow at Emmanuel College, Cambridge.


\begin{thebibliography}{10}

\bibitem{Kampen81a}
N.~G. van Kampen, {\em Stochastic Processes in Physics and Chemistry}
  (North-Holland, Amsterdam, 1981).

\bibitem{Stillinger82a}
F.~H. Stillinger and T.~A. Weber, Phys. Rev. A {\bf 25},  978  (1982).

\bibitem{Wales98c}
D.~J. Wales, M.~A. Miller, and T.~R. Walsh, Nature {\bf 394},  758  (1998).

\bibitem{Miller99a}
M.~A. Miller, J.~P.~K. Doye, and D.~J. Wales, J. Chem. Phys. {\bf 110},  328
  (1999).

\bibitem{LJpaper}
J.~P.~K. Doye, M.~A. Miller, and D.~J. Wales, (cond-mat/9903305).

\bibitem{p46}
M.~A. Miller and D.~J. Wales, (cond-mat/9904304).

\bibitem{Kunz98a}
R.~E. Kunz, P. Blaudeck, K.~H. Hoffmann, and R.~S. Berry, J. Chem. Phys. {\bf
  108},  2576  (1998).

\bibitem{Czerminski90a}
R. Czerminski and R. Elber, J. Chem. Phys. {\bf 92},  5580  (1990).

\bibitem{Press92a}
W.~H. Press, S.~A. Teukolsky, W.~T. Vetterling, and B.~P. Flannery, {\em
  Numerical Recipes in FORTRAN}, 2 ed. (Cambridge University Press, Cambridge,
  1992).

\bibitem{Vekhter97a}
B. Vekhter and R.~S. Berry, J. Chem. Phys. {\bf 106},  6456  (1997).

\bibitem{Miller97a}
M.~A. Miller and D.~J. Wales, J. Chem. Phys. {\bf 107},  8568  (1997).

\bibitem{Gilbert90a}
R.~G. Gilbert and S.~C. Smith, {\em Theory of Unimolecular and Recombination
  Reactions} (Blackwell, Oxford, 1990).

\bibitem{Hoare79a}
M.~R. Hoare, Adv. Chem. Phys. {\bf 40},  49  (1979).

\bibitem{frankehb93}
G. Franke, E.~R. Hilf, and P. Borrmann, J. Chem. Phys. {\bf 98},  3496  (1993).

\bibitem{Wales93f}
D.~J. Wales, Mol. Phys. {\bf 78},  151  (1993).

\bibitem{Morse29a}
P.~M. Morse, Phys. Rev. {\bf 34},  57  (1929).

\bibitem{Girifalco59a}
L.~A. Girifalco and V.~G. Weizer, Phys. Rev. {\bf 114},  687  (1959).

\bibitem{Girifalco92a}
L.~A. Girifalco, J. Phys. Chem. {\bf 96},  858  (1992).

\bibitem{Wales94c}
D.~J. Wales and J. Uppenbrink, \prb {\bf 50},  342  (1994).

\bibitem{Ball98a}
K.~D. Ball and R.~S. Berry, J. Chem. Phys. {\bf 109},  8541  (1998).

\bibitem{Ball98b}
K.~D. Ball and R.~S. Berry, J. Chem. Phys. {\bf 109},  8557  (1998).

\bibitem{Haarhoff63a}
P.~C. Haarhoff, Mol. Phys. {\bf 7},  101  (1963).

\bibitem{Doye95a}
J.~P.~K. Doye and D.~J. Wales, J. Chem. Phys. {\bf 102},  9659  (1995).

\bibitem{Rose93b}
J.~P. Rose and R.~S. Berry, J. Chem. Phys. {\bf 98},  3262  (1993).

\bibitem{Socci96a}
N.~D. Socci, J.~N. Onuchic, and P.~G. Wolynes, J. Chem. Phys. {\bf 104},  5860
  (1996).

\bibitem{Doye96b}
J.~P.~K. Doye and D.~J. Wales, J. Chem. Phys. {\bf 105},  8428  (1996).

\bibitem{Socci94a}
N.~D. Socci and J.~N. Onuchic, J. Chem. Phys. {\bf 101},  1519  (1994).

\bibitem{Stillinger95a}
F.~H. Stillinger, Science {\bf 267},  1935  (1995).

\bibitem{Koper87a}
G.~J.~M. Koper and H.~J. Hilhorst, Europhys. Lett. {\bf 3},  1213  (1987).

\bibitem{Palmer84a}
R.~G. Palmer, D.~L. Stein, E. Abrahams, and P.~W. Anderson, Phys. Rev. Lett.
  {\bf 53},  958  (1984).

\bibitem{Skorobogatiy98a}
M. Skorobogatiy, H. Guo, and M. Zuckermann, J. Chem. Phys. {\bf 109},  2528
  (1998).

\bibitem{Scherer92a}
G.~W. Scherer, J. Am. Ceram. Soc. {\bf 75},  1060  (1992).

\bibitem{Angell91a}
C.~A. Angell, J. Non-cryst. Solids {\bf 131--133},  13  (1991).

\bibitem{Golub89a}
G.~H. Golub and C.~F. van Loan, {\em Matrix Computations}, 2nd ed. (John
  Hopkins University Press, Baltimore, 1989).

\bibitem{Jones25a}
J.~E. Jones and A.~E. Ingham, Proc. R. Soc. A {\bf 107},  636  (1925).

\bibitem{Doye99b}
J.~P.~K. Doye, M.~A. Miller, and D.~J. Wales, J. Chem. Phys. {\bf 110},  6896
  (1999).

\bibitem{Doye98b}
J.~P.~K. Doye, D.~J. Wales, and M.~A. Miller, J. Chem. Phys. {\bf 109},  8143
  (1998).

\bibitem{Kunz95a}
R.~E. Kunz and R.~S. Berry, J. Chem. Phys. {\bf 103},  1904  (1995).

\bibitem{Chandler}
D. Chandler, {\em Introduction to Modern Statistical Mechanics} (Oxford
  University Press, Oxford, 1987).

\bibitem{Angelani98a}
L. Angelani, G. Parisi, G. Ruocco, and G. Viliani, Phys. Rev. Lett. {\bf 81},
  4648  (1998).

\bibitem{Angelani99a}
L. Angelani, G. Parisi, G. Ruocco, and G. Viliani, (cond-mat/9904125).

\bibitem{Sibani93a}
P. Sibani, J.~C. Sch\"on, P. Salamon, and J. Andersson, Europhys. Lett. {\bf
  22},  479  (1993).

\bibitem{Schon96a}
J.~C. Sch\"on, Ber. Bunsenges. Phys. Chem. {\bf 100},  1388  (1996).

\bibitem{Schon96b}
J.~C. Sch\"on, H. Putz, and M. Jansen, J. Phys. Condensed Matter {\bf 8},  143
  (1996).

\bibitem{Kiefhaber95a}
T. Kiefhaber, Proc. Natl. Acad. Sci. USA {\bf 92},  9029  (1995).

\bibitem{Baker94a}
D. Baker and D.~A. Agard, Biochemistry {\bf 33},  7505  (1994).

\bibitem{Jackson99a}
G.~S. Jackson {\it et~al.}, Science {\bf 283},  1935  (1999).

\end{thebibliography}



\end{multicols}

\end{document}